\documentclass[onecolumn,superscriptaddress,amsmath,amssymb,aps,showkeys,floatfix]{revtex4-2}
\usepackage{amsmath,bm}
\usepackage{amssymb}
\usepackage{amsthm}
\usepackage{amsfonts}

\usepackage{mathtools}
\usepackage{rotating}
\usepackage{capt-of}
\usepackage{array}
\usepackage{blkarray, bigstrut}
\usepackage{enumitem}
\newlist{notes}{enumerate}{1}
\setlist[notes]{label=Note: ,leftmargin=*}
\usepackage{graphicx}
\usepackage{hyperref}
\usepackage{xcolor}
\usepackage{comment}
\usepackage{makecell}

\usepackage{capt-of}
\usepackage{float}
\makeatletter
\let\newfloat\newfloat@ltx
\makeatother

\usepackage{algorithm}
\usepackage{algpseudocode}
\usepackage{float}
\usepackage{enumitem}
\usepackage[section]{placeins}
\usepackage{bm}% bold mathhttps://www.overleaf.com/project/63898e941dc3f81d3b516496
\usepackage[normalem]{ulem}
\theoremstyle{definition}

\usepackage{multirow}
\usepackage{booktabs}
\usepackage{rotating,tabularx}
\usepackage{ulem}
\usepackage{url}
\usepackage{hhline}
\usepackage{siunitx}
\usepackage{dcolumn}% Align table columns on decimal point
\renewcommand{\thetable}{\arabic{table}}
\newcolumntype{T}[1]{>{\centering\arraybackslash\hspace{0pt}}p{#1}}
\newcolumntype{P}[1]{>{\centering\arraybackslash}p{#1}}

\makeatletter
\def\maketitle{
\@author@finish
\title@column\titleblock@produce
\suppressfloats[t]}
\makeatother

\renewcommand{\thesection}{\arabic{section}}

\makeatletter
\def\p@subsection{}
\makeatother
\makeatletter
\def\p@subsubsection{}
\makeatother
\begin{document}
\preprint{APS}
%====================================================

%=================================================
% Title
\title{A comparative exploration of Steinerberger curvature and resistance curvature in model and real-world complex networks}
%================================================

%===================================================================
% Authors

\author{Madhumita Mondal}
\affiliation{The Institute of Mathematical Sciences (IMSc), Chennai 600113, India}
\affiliation{Homi Bhabha National Institute (HBNI), Mumbai 400094, India}

\author{Priyank Vasu}
\affiliation{The Institute of Mathematical Sciences (IMSc), Chennai 600113, India}

\author{Areejit Samal}
\thanks{To whom correspondence should be addressed:\\ asamal@imsc.res.in or jjost@mis.mpg.de}
\affiliation{The Institute of Mathematical Sciences (IMSc), Chennai 600113, India}
\affiliation{Homi Bhabha National Institute (HBNI), Mumbai 400094, India}

\author{J\"urgen Jost}
\thanks{To whom correspondence should be addressed:\\ asamal@imsc.res.in or jjost@mis.mpg.de}
\affiliation{Max Planck Institute for Mathematics in the Sciences, 04103 Leipzig, Germany}
\affiliation{Center for Scalable Data Analytics and Artificial Intelligence Dresden/Leipzig, Leipzig, Saxony, 04107, Germany}
\affiliation{Santa Fe Institute for the Sciences of Complexity, Santa Fe, New Mexico 87501, USA}

%===============================================================

%===================================================================
% Abstract

\begin{abstract}
Discrete curvature extends the concept of curvature from continuous spaces to discrete structures such as graphs and networks. 
It provides a mathematical framework for characterising local and global network geometry, that offers insights into structural properties such as connectivity and clustering. 
In this study, we investigate Steinerberger curvature (SBC) and resistance curvature (RC) and examine their relationships with established graph curvature measures, namely Forman-Ricci, Augmented Forman-Ricci, and Ollivier–Ricci curvatures, a notion of scalar sectional curvature, as well as conventional vertex-based measures. 
A systematic comparative analysis is conducted across various synthetic network models and real-world networks, focusing on curvature distributions, correlations with structural measures, and the identification of structurally important vertices through network disintegration experiments based on global efficiency. 
Our results show that SBC generally exhibits weaker correlations with other measures. 
On the other hand, RC is strongly positively correlated with established discrete Ricci curvature measures, whereas it is predominantly negatively correlated with conventional vertex-based measures like degree and centrality-based measures. 
Furthermore, RC achieves a reduction in communication efficiency comparable to or faster than established Ricci curvature measures. 
These findings suggest that SBC captures structural characteristics that are largely distinct from those reflected by other measures, while RC exhibits distinct structural associations and shows significant potential for identifying vertices critical to network connectivity.
\end{abstract}

\maketitle

%=====================================================
% Introduction
\section{Introduction}

Curvature is a fundamental concept in differential geometry that describes the intrinsic and extrinsic geometric properties of spaces and their deviation from flat Euclidean geometry. In the classical setting, several notions of curvature have been developed to capture different aspects of geometry, including Gaussian curvature, mean curvature, sectional curvature, Ricci curvature, and scalar curvature \cite{jost2008riemannian,gallot2004riemannian, sandhu2015graph}. With the development of discrete and combinatorial approaches to geometry, researchers have extended the concept of curvature beyond smooth manifolds to discrete structures such as graphs, simplicial complexes, and CW complexes \cite{whitehead1949combinatorial}. These extensions aim to capture geometric information from combinatorial, metric, and topological properties without necessarily requiring an underlying smooth structure. Consequently, a variety of discrete curvature notions have emerged, each motivated by a different aspect of classical curvature or by a distinct mathematical framework. For example, Forman's discretization is based on a combinatorial analogue of the classical Bochner Weitzenböck formula in differential geometry \cite{forman2003bochner}, Ollivier's coarse Ricci curvature for metric spaces equipped with a measure or random walk \cite{ollivier2007ricci, ollivier2009ricci}, and Bakry-\'Emery's notion originated from a so called Bochner identity \cite{gallot2004riemannian}, as well as a notion of sectional curvature that provides a global perspective on geometry \cite{joharinad2019topology}. Also, a notion of resistance curvature was introduced by Devriendt and Lambiotte, based on effective resistance inspired by electrical circuit theory in \cite{devriendt2022discrete}.

First, these notions have been investigated primarily from a theoretical perspective and have later been applied to the analysis of complex networks and other discrete structures \cite{sreejith2016forman, samal2018comparative, saucan2019discrete, mondal2024bakry}. Network science investigates the structural properties and organizational principles of networks composed of interconnected nodes and edges, with the aim of identifying common patterns as well as features that distinguish different classes of networks \cite{networkscience2016}. Networks provide a natural framework for representing complex systems arising in diverse fields, including biology, transportation, communication, and social sciences \cite{sia2019ollivier, samal2021network, elumalai2022graph, yadav2023discrete}. To characterize the structure and organization of such systems, researchers have increasingly employed geometry-inspired measures that capture different aspects of network structure \cite{sreejith2016forman,samal2018comparative,cushing2020bakry,elumalai2022graph,yadav2023discrete,fesser2023augmentations}. 

In particular, discrete notions of curvature have attracted considerable attention as tools for investigating the local and global structural properties of complex networks \cite{sreejith2016forman, samal2018comparative, saucan2019discrete,mondal2024bakry,sia2019ollivier}. By assigning curvature values commonly to vertices or edges, these approaches provide geometric perspectives on network structure and can help identify features such as bottlenecks, clustering, and community detection. Consequently, researchers have studied and applied several discrete curvature notions in network analysis, although their mathematical foundations and the structural characteristics they capture vary by definition; for details, see \cite{YadavXia2025}. 

In this study, we investigate Steinerberger curvature \cite{Steinerberger2023} and resistance curvature. 
The notion of resistance curvature was initially introduced by Devriendt and Lambiotte \cite{devriendt2022discrete}. 
Subsequently, building on that work and connecting it to Steinerberger curvature \cite{Steinerberger2023}, Devriendt, Ottolini, and Steinerberger \cite{devriendt2024graph} proposed a related formulation based on the effective resistance matrix.
In the present work, we adopt the latter formulation and refer to it simply as \textit{resistance curvature} throughout the article.
Both notions provide ways to characterize the structural properties of networks through distance-based information. Unlike conventional graph measures, such as degree and centrality, which primarily describe local connectivity or a vertex's position within the network, these curvatures incorporate global information \cite{Steinerberger2023,devriendt2024graph}. This makes them potentially useful tools for investigating structural patterns across different network topologies.
We examine their behavior across different synthetic network models and real-world networks to understand how they vary with network topology. We also investigate their relationships with commonly used graph measures and established discrete Ricci curvature measures \cite{networkscience2016}, including Forman–Ricci \cite{forman2003bochner, samal2018comparative} and Ollivier–Ricci curvature \cite{ollivier2007ricci}. Additionally, we compare Steinerberger curvature and resistance curvature with scalar sectional curvature to identify similarities and differences in the structural information captured by these measures. Through these comparisons, we aim to determine whether Steinerberger curvature and resistance curvature capture structural characteristics that are not fully reflected by existing measures and to assess their potential as complementary tools for network analysis. 
It is worth clarifying that the resistance curvature introduced by Devriendt and Lambiotte \cite{devriendt2022discrete} and the curvature defined by Devriendt, Ottolini, and Steinerberger \cite{devriendt2024graph} are essentially equivalent, differing only by a scalar factor at each vertex. The resistance curvature introduced by Devriendt and Lambiotte has also been used to establish theoretical relationships with Forman–Ricci and Ollivier–Ricci curvatures. However, subsequent studies have primarily focused on specific applications, such as graph neural networks (GNNs), hyperuniformity, and particular classes of graphs \cite{raj2025local,estrada2025forman,chen2026rethinking, fei2025efficient}. Here, we undertake a systematic comparative study of these curvature measures across a range of synthetic network models and real-world networks, with the aim of providing a comprehensive understanding of their behaviour.

The subsequent sections are organized as follows. 
Section \ref{sec:theory} presents the theoretical background, including the definitions of Steinerberger curvature, resistance curvature, Forman-Ricci curvature, Ollivier-Ricci curvature, and scalar sectional curvature. 
Section \ref{ses:dataset} describes the model networks and real-world networks considered in the analysis, along with the network construction procedures. 
Section \ref{sec:results} presents the results and discusses the main findings, focusing on the distribution of Steinerberger and resistance curvature and its relationships with other curvature measures and conventional graph measures. 
Finally, Section \ref{sec:conclusion} concludes the paper by highlighting the main findings, their implications, and future directions of the study.

%=====================================================

%=====================================================
\section{Theory}\label{sec:theory}
In Riemannian geometry, curvature is a local invariant of a metric space that measures its deviation from Euclidean geometry. It is most generally described by the Riemann curvature tensor, which captures the non-commutativity of covariant derivatives on a manifold \cite{stone1976combinatorial, jost1997nonpositive,  chow2003combinatorial, gallot2004riemannian, morgan2005manifolds, lott2009ricci, jost2008riemannian, joharinad2019topology}. 
Important contractions and restrictions of the Riemann curvature tensor give rise to several fundamental notions, including sectional, Ricci, and scalar curvature. 
In the discrete setting, various notions of curvature have been introduced to capture analogous geometric properties of graphs. 
For details, one can refer to the review article by Yadav and Xia \cite{YadavXia2025}. 
In this section, we briefly review five curvature notions which can be defined on the vertices of graphs and explored in this empirical study.

%=============================================================
\subsection{Steinerberger Curvature on Graphs}

Steinerberger curvature is a theoretical notion of curvature defined on a finite connected graph through an equilibrium measure for the graph distance (see \cite{Steinerberger2023}).
Let $G=(V, E)$ be a finite connected graph with $|V|=n$, and let $d(u,v)$ denote the graph distance between two vertices $u,v\in V$.
Steinerberger introduced a notion of curvature by considering a signed measure $\mu$ on $V$ whose distance potential is constant, namely
\begin{equation}
    \sum_{v\in V} d(u,v)\mu(v)=n,
    \qquad u\in V.
    \label{eq:equilibrium_measure}
\end{equation}
The quantity $\mu(v)$ is interpreted as the curvature at the vertex
$v$. Enumerating $V=\{v_1,\ldots,v_n\}$ and defining the graph distance
matrix
\begin{equation*}
    D=\bigl(d_G(v_i,v_j)\bigr)_{i,j=1}^{n},
\end{equation*}
together with the curvature vector $\kappa=  \bigl(\kappa(v_1),\ldots,\kappa(v_n)\bigr)^T,$
equation \eqref{eq:equilibrium_measure} can be equivalently written as
\begin{equation}\label{eq:steinerberger_curvature}
    D\kappa=n\mathbf{1},
\end{equation}
where $\mathbf{1}$ denotes the all ones vector. We will use the entries $\kappa_i$ of the curvature vector $\kappa$ interchangeably if the vertices $V$ are enumerated. 
The distance matrix $D$ is generally computed using Dijkstra’s algorithm \cite{Dijkstra59}, with a time complexity of $\mathcal{O}(m + n\log n)$, where $|E|=m$.
The solvability of the curvature equation \eqref{eq:steinerberger_curvature} depends on the algebraic properties of the distance matrix $D$, which are described below.

\begin{enumerate}
    \item \textbf{Unique solution:} If $D$ is invertible, i.e., $\det(D)\neq0$, then the system has a unique solution given by
    \begin{equation*}
        \kappa=nD^{-1}\mathbf{1}.
        \label{eq:inverse_curvature}
    \end{equation*}

    \item \textbf{Multiple solutions:} If $D$ is singular but $n\mathbf{1}\in\operatorname{Im}(D)$, then the system is consistent and admits, in general, multiple exact solutions. If $\kappa_{o}$ is one such solution, then the complete solution set is
    \begin{equation*}
        \kappa=\kappa_{o}+z,
        \qquad z\in\ker(D).
        \label{eq:affine_solution}
    \end{equation*}
     Now, a canonical choice is obtained by maximising the lower curvature bound $K=\min_{1\leq i\leq n}\kappa_i.$

    \item \textbf{No exact solution:} If $D$ is singular and $n\mathbf{1}\notin\operatorname{Im}(D)$, then the system has no exact solution. In this case, the Moore–Penrose pseudoinverse \cite{penrose1955generalized, barata2012moore} of $D$ says $D^\dagger$ provides the solution as
    \begin{equation*}
        \kappa=D^\dagger(n\mathbf{1}).
        \label{eq:pseudoinverse_curvature}
    \end{equation*}
\end{enumerate}

Steinerberger curvature is determined by the global metric geometry of the graph through its distance matrix. 
In particular, it is a non-local curvature notion since a modification of the graph may change the shortest path distances between many pairs of vertices, and hence, alter the curvature vector.
Moreover, the total curvature is defined by
\begin{equation}
    \mathcal{K}(G)
    :=
    \sum_{v\in V}\kappa(v).
    \label{eq:total_curvature}
\end{equation}
If $\kappa^1,\kappa^2\in\mathbb{R}^{n}$ are two solutions of
\eqref{eq:steinerberger_curvature}, then the total curvature 
\begin{equation*}
    \label{eq:total_curvature_invariance}
    \mathcal{K}(G)
    =
    \sum_{i=1}^n\kappa^1_i=
    \sum_{i=1}^n\kappa^2_i,
\end{equation*}
is an invariant \cite{ChenTsui2025}.

Steinerberger curvature theory has strong geometric consequences, including Bonnet–Myers, Cheng, Lichnerowicz, reverse Bonnet–Myers, and minimax results (see \cite{Steinerberger2023}). Thus, Steinerberger curvature captures both the global metric and spectral geometry of a graph, in analogy with classical curvature bounds \cite{Steinerberger2023, CushingEtAl2025}. Moreover, it is intrinsic, and its computation reduces to solving the linear system $D\kappa=n\mathbf{1}$, making it comparatively simple to compute and allowing for explicit calculations for several important graph families.

%=============================================================

\subsection{Resistance Curvature}
The first version of a global curvature of a graph was initially proposed by Devriendt and Lambiotte \cite{devriendt2022discrete}. 
Their notion of discrete curvature is based on effective resistance.
Effective resistance provides an alternative measure of distance between vertices of a graph, incorporating information from all available paths rather than considering only the shortest path. Building on this property, Devriendt, Ottolini, and Steinerberger \cite{devriendt2024graph} observed that a different normalization of that resistance curvature makes a framework possible that includes both Steinerberger and effective resistance curvature.
This approach is particularly useful because effective resistance is closely connected to electrical network theory and random walks, offering a way to relate curvature to the structural and spectral properties of graphs. 
The effective resistance between two vertices of a graph depends not only on the shortest connection between them, that is, their distance, but also on other paths connecting them, in analogy to the effective resistance of an electrical network.

Let \(G=(V,E)\) be a connected graph with \(n\) vertices. Let \(A\) denote its adjacency matrix, \(\mathcal{D}\) its diagonal degree matrix, and \(L=\mathcal{D}-A\) its Kirchhoff Laplacian. If \(\mathbf{1}\in\mathbb{R}^n\) denotes the all ones vector, define
\[
\Gamma=L+\frac{1}{n}\mathbf{1}\mathbf{1}^{\mathsf T}.
\]
Since \(G\) is connected, \(\Gamma\) is invertible. The effective resistance between vertices \(v_i\) and \(v_j\) is given by
\[
\Omega_{ij}
=(\Gamma^{-1})_{ii}+(\Gamma^{-1})_{jj}
-2(\Gamma^{-1})_{ij}.
\]
The resulting matrix \(\Omega=(\Omega_{ij})\) is called the effective resistance matrix of \(G\).

The resistance curvature at a vertex \(v_i\in V\) is defined as the \(i\)-th component of the vector \(\kappa_R=(\kappa_1,\ldots,\kappa_n)^{\mathsf T}\) satisfying
\[
\Omega\kappa_R=\mathbf{1}.
\]
For every finite connected graph, the effective resistance matrix \(\Omega\) is invertible. Consequently, the system has a unique solution, given by \(\kappa_R=\Omega^{-1}\mathbf{1}\); see \cite{devriendt2024graph}.
This definition is closely related to the normalized resistance curvature introduced by Devriendt and Lambiotte \cite{devriendt2022discrete}, which can be expressed as
\[
p=
\frac{\Omega^{-1}\mathbf{1}}
{\langle\mathbf{1},\Omega^{-1}\mathbf{1}\rangle}.
\]
Thus, the normalized curvature vector is obtained by dividing \(\kappa_R\) by the sum of its components. Equivalently,
$
\kappa_R=
\langle\mathbf{1},\Omega^{-1}\mathbf{1}\rangle\,p.
$
Hence, the two curvature vectors differ only by a global normalization factor, which preserves their relative proportions but may change the magnitudes of the individual curvature values.\\
We also point out that the matrix $\Omega$ is always invertible. Therefore, in contrast to the Steinerberger curvature, the resistance curvature is always uniquely defined.

%=============================================================

%=============================================================
\subsection{Forman-Ricci Curvature}

Forman introduced a combinatorial notion of Ricci curvature for weighted CW-complexes based on a combinatorial analogue of the classical Bochner-Weitzenbo\"ck formula in differential geometry, which relates curvature to the Laplace operator and provides a discrete analogue of the corresponding relation in Riemannian geometry; for details, see \cite{whitehead1949combinatorial,forman2003bochner,jost2008riemannian}. 
Subsequently, the Forman–Ricci curvature framework has been adapted for the analysis of graphs; for details see \cite{sreejith2016forman,samal2018comparative,saucan2018discrete,saucan2019discrete}. 
For an unweighted graph ${G}=({V},{E})$, the Forman-Ricci curvature of an edge
$e\in{E}$ is given by
\begin{equation}
    F(e)=4-\sum_{v\sim e}\deg(v),
    \label{FRCunweighted}
\end{equation}
where $v\sim e$ indicates that $v$ is an endpoint of the edge $e$. 

We also consider the augmented Forman-Ricci curvature introduced in
\cite{samal2018comparative}. This extension incorporates additional local
connectivity information through the clique complex, in which complete
subgraphs are represented as simplices. In particular, triangles containing
an edge contribute explicitly to the augmented curvature. Thus, the
augmented formulation takes triangular structures in a graph into account, while cycles of length greater than three do not contribute directly.
For an unweighted graph ${G}=({V},{E})$, the augmented ($F^{\#}(e)$) and standard ($F(e)$) Forman-Ricci
curvatures are related by
\begin{equation}
    F^{\#}(e)=F(e)+3m,
    \label{AFRC}
\end{equation}
where $m$ denotes the number of triangles containing the edge $e$
\cite{samal2018comparative}. 

Although Forman-Ricci curvature is naturally an edge-based quantity, a vertex-level quantity can also be obtained by aggregating the curvatures of the edges incident to the vertex \cite{samal2018comparative,sreejith2017systematic}. 
In particular, for $v\in{V}$, we define the vertex-based Forman-Ricci curvature by
\begin{equation}
    F(v)
    =
    \sum_{e\in E(v)} F(e),
    \label{vertex_FRC}
\end{equation}
where $E(v)$ is the set of edges
incident to $v$, and $F(e)$ is the Forman-Ricci curvature of the edge $e$.
Similarly, the vertex-based augmented Forman-Ricci curvature is defined as
\begin{equation}
    {F^{\#}}(v)
    =
    \sum_{e\in E(v)} F^{\#}(e).
    \label{vertex_AFRC}
\end{equation}
Thus, the edge-based Forman-Ricci and augmented Forman-Ricci curvatures can be represented at the vertex level, allowing their comparison with other curvature notions defined on graphs. 
This is analogous to the definition of scalar curvature in Riemannian geometry (see \cite{jost2008riemannian}), which is a point-based quantity obtained by summing the Ricci curvatures for the tangent directions emanating from the point in question. Therefore, we might also call the preceding the \textit{(augmented) Forman scalar curvature}.
%=============================================================

%=============================================================
\subsection{Ollivier-Ricci Curvature}

Ollivier introduced the notion of coarse Ricci curvature for metric spaces equipped with a probability measure or a random walk \cite{ollivier2007ricci,ollivier2009ricci}. Since then, it has been extensively developed for discrete spaces, especially graphs and complex networks; for details, see \cite{lin2010ricci,bauer2012ollivier,samal2018comparative,elumalai2022graph, yadav2023discrete}. Its broad use in network geometry also makes Ollivier-Ricci curvature an important reference for comparing different notions of discrete curvature. The curvature is defined as follows.

Let ${G}=({V},{E})$ be a graph with vertex set $V$ and edge set $E$. The construction of Ollivier-Ricci curvature begins by assigning a probability measure to each vertex and comparing the measures associated with pairs of vertices using the Wasserstein distance. Let $v_i\in{V}$ be a vertex of degree $k$, and let $N_{v_i}$ denote its set of neighbors. For $\alpha\in[0,1]$, the probability measure $m_i^\alpha$ associated with $v_i$ is defined by
\[
m_i^{\alpha}(v)=
\begin{cases}
\alpha, & \text{if } v=v_i,\\[2mm]
\dfrac{1-\alpha}{k}, & \text{if } v\in N_{v_i},\\[2mm]
0, & \text{otherwise}.
\end{cases}
\]
Thus, the measure assigns mass $\alpha$ to $v_i$ and distributes the remaining mass $1-\alpha$ uniformly among its neighbors.
Let $d(v_i,v_j)$ denote the graph distance between two vertices $v_i,v_j\in V$. The
Ollivier-Ricci curvature of an edge $e=(v_i,v_j)\in{E}$ is defined
as
\[
\kappa_O(e)
=
1-\frac{W_1(m_i^\alpha,m_j^\alpha)}
{d(v_i,v_j)},
\]
where $W_1$ denotes the $1$-Wasserstein distance between the two probability
measures. It is given by
\[
W_1(m_i^\alpha,m_j^\alpha)
=
\inf_{\mu_{ij}\in\Pi(m_i^\alpha,m_j^\alpha)}
\sum_{(v_k,v_l)\in{V}\times{V}}
d(v_k,v_l)\mu_{ij}(v_k,v_l),
\]
where $\Pi(m_i^\alpha,m_j^\alpha)$ denotes the set of all couplings of
$m_i^\alpha$ and $m_j^\alpha$. Thus, for
$\mu_{ij}\in\Pi(m_i^\alpha,m_j^\alpha)$, we have
\[
\sum_{v_l\in{V}}
\mu_{ij}(v_k,v_l)
=
m_i^\alpha(v_k),
\qquad
\sum_{v_k\in{V}}
\mu_{ij}(v_k,v_l)
=
m_j^\alpha(v_l).
\]
Therefore, $W_1(m_i^\alpha,m_j^\alpha)$ represents the minimum cost required to transport the probability measure centered at $v_i$ to that centered at $v_j$, where the transportation cost is measured using the graph distance. Thus, Ollivier-Ricci curvature compares the optimal transportation cost with the distance between the corresponding vertices. 

Although Ollivier-Ricci curvature is naturally defined on edges, a vertex-based quantity, that is, an \textit{Ollivier scalar curvature}, can also be obtained by summing the curvatures of the edges incident to that vertex \cite{samal2018comparative}. In particular, for
$v\in{V}$, we define
\[
\kappa_O(v)
=
\sum_{e\in E(v)}\kappa_O(e),
\]
where $E(v)$ is the set of edges
incident to $v$, and $\kappa_O(e)$ denotes the Ollivier-Ricci curvature of
the edge $e$.

%=============================================================

%=============================================================
\subsection{Sectional Curvature}

Joharinad and Jost \cite{JoharinadJost2019,JoharinadJost2023}
introduced a notion of sectional curvature for general metric spaces. Their construction is motivated by Gromov's metric characterizations of non-positive curvature. In contrast to classical curvature notions, which characterize the departure of a metric from Euclidean geometry, this construction measures how a metric space deviates from the reference geometry of a tripod. A geodesic length space $(X,d)$ is called a tripod space if, for every three points $x_1,x_2,x_3\in X$, there exists a point
$m\in X$, called a \textit{median}, such that
\[
d(x_i,m)+d(x_j,m)=d(x_i,x_j),
\qquad 1\leq i<j\leq 3.
\]
Thus, the three geodesic segments joining the points $x_1,x_2,x_3$ meet at
the common point $m$.

The tripod property can be expressed through the intersection property for closed balls. Namely, for any $x_1,x_2,x_3\in X$ and $r_1,r_2,r_3>0$ satisfying
\[
r_i+r_j\geq d(x_i,x_j),
\qquad 1\leq i<j\leq 3,
\]
one has
\[
\bigcap_{i=1}^{3}B(x_i,r_i)\neq\varnothing,
\]
where
\[
B(x_i,r_i):=\{x\in X:d(x_i,x)\leq r_i\}.
\]
This intersection property characterizes the tripod geometry and serves as the basis for quantifying deviations from it in general metric spaces.

To define the sectional curvature associated with a triple of points
$x_1,x_2,x_3$, consider three closed balls
$B(x_1,r_1)$, $B(x_2,r_2)$, and $B(x_3,r_3)$ whose radii are chosen so that
each pair of balls intersects precisely at the boundary. Thus,
\[
r_i+r_j=d(x_i,x_j),\qquad 1\leq i<j\leq3.
\]
The resulting system has the unique solution
\begin{align*}
r_1 &=
\frac{1}{2}\bigl[d(x_1,x_2)+d(x_1,x_3)-d(x_2,x_3)\bigr],\\
r_2 &=
\frac{1}{2}\bigl[d(x_1,x_2)+d(x_2,x_3)-d(x_1,x_3)\bigr],\\
r_3 &=
\frac{1}{2}\bigl[d(x_1,x_3)+d(x_2,x_3)-d(x_1,x_2)\bigr].
\end{align*}
These quantities are the corresponding Gromov products associated with the
three-point configuration.

Starting from these radii, the three balls are enlarged proportionally until
they acquire a common intersection. More precisely, the expansion constant
$\rho(x_1,x_2,x_3)$ is defined as the infimum of all $\mu$ for which
\[
\bigcap_{i=1}^{3}B(x_i,\mu r_i)\neq\varnothing.
\]
Equivalently, the sectional curvature associated with the triple can be
written as
\begin{equation}\label{sectional curvature for metric space}
\rho(x_1,x_2,x_3)
:=
\inf_{x\in X}
\max_{i=1,2,3}
\frac{d(x_i,x)}{r_i}. 
\end{equation}
For complete metric spaces, one has
\[
1\leq \rho(x_1,x_2,x_3)\leq2.
\]
The lower bound is attained in tripod spaces. Indeed, whenever the three balls intersect pairwise, their common intersection is nonempty; in the boundary case, the median $m$ belongs to this intersection. Thus, $\rho$ quantifies the deviation of the given three-point configuration from the tripod property. Larger values of $\rho$ indicate a greater enlargement of the balls required to obtain a common intersection. Thus, $\rho$ is intrinsically a three-point quantity, assigning a value in
$[1,2]$ to each triple of points. One then compares that value with the corresponding value for a comparison triangle, that is, a triangle with the same pairwise distances between its three vertices, in a model space. For instance, when the value of $\rho$ is never larger than the value in such a comparison triangle in the Euclidean plane, we say that the space has non-positive curvature.
Further details can be found in \cite{JoharinadJost2019,JoharinadJost2023,BeylierEtAl2025,YadavXia2025}.

For computational purposes, we adopt the approach of \cite{BeylierEtAl2025}, which restricts the analysis to triangles of a prescribed shape, in particular, equilateral triangles. Suppose that
the three vertices have the same pairwise distance $d>0$.
Then $r_1=r_2=r_3=d/2$. Placing a ball of initial radius
$r_{\mathrm{in}}=d/2$ at each vertex and increasing all three
radii simultaneously, we determine the infimum
$r_{\mathrm{out}}$ at which the balls have a common
intersection. The corresponding expansion factor is
\begin{equation*}
\rho(x_1,x_2,x_3)=\frac{r_{\mathrm{out}}}{r_{\mathrm{in}}}.
\end{equation*}
This procedure provides a direct computational method for estimating the sectional curvature associated with a collection of equilateral configurations, without explicitly solving the minimization problem defining $\rho$ in equation \eqref{sectional curvature for metric space}.

It is also important to note that graphs naturally possess a metric structure. For an unweighted graph ${G}=({V},{E})$, the distance between two vertices is defined as the length of the shortest path connecting them. Thus, the graph distance induces a metric on the vertex set ${V}$. Consequently, the above metric space construction can be applied directly to graphs by considering triples of vertices and their pairwise graph distances. 
This provides an intrinsic curvature-based framework for studying the geometric properties of networks.
Although sectional curvature is fundamentally defined on triples of vertices, Yadav and Xia \cite{YadavXia2025} provide a way to extend this construction to obtain a vertex-based formulation by averaging the sectional curvature over suitable collections of vertex triples; in our case, we consider equilateral triples.
Let $T_v$ denote the collection of equilateral triples containing a vertex $v\in V$; if $T_v$ is non-empty, define the corresponding vertex-based quantity as
\[
\kappa(v)
=
\frac{1}{|T_v|}
\sum_{\{v_1,v_2,v_3\}\in T_v}
\rho(v_1,v_2,v_3).
\]
These aggregation procedures provide a vertex-level representation of the inherently three-point sectional curvature, yielding a discrete analogue of scalar curvature and facilitating comparisons with other curvature constructions in network geometry. 
Throughout this paper, we refer to this vertex-based quantity simply as \textit{scalar sectional curvature}.

%=============================================================

\section{Dataset of Model and Real-World Networks}\label{ses:dataset}
\label{Datasets}
%=================================================================

To evaluate the different curvature notions considered in this work, we use both model networks and real-world networks. All networks considered in the
analysis are undirected and unweighted. 
The model networks are generated using four standard random graph models: Erd\"os-R\'enyi (ER), Watts-Strogatz (WS), Barab\'asi-Albert (BA), and the Hyperbolic Graph Generator (HGG).

\begin{itemize}
\item \textbf{Erd\"os-R\'enyi (ER) model} \cite{engel2004large}:
An ER network is denoted by $G(n,p_{_{ER}})$, where $n$ is the number of vertices and $p_{_{ER}}$ is the probability that a pair of vertices is connected by an edge. 
Each possible edge is generated independently with the same probability $p_{_{ER}}$.

\item \textbf{Watts-Strogatz (WS) model} \cite{watts1998collective}:
The WS model produces networks with small-world characteristics, namely, relatively high clustering together with short characteristic path lengths. 
The construction starts with a regular ring lattice containing $n$ vertices, where each vertex is connected to its $k_{_{WS}}$ nearest neighbors. 
Each edge is then considered for rewiring with probability $p_{_{WS}}$, and one endpoint is connected to another vertex chosen uniformly at random, subject to the avoidance of self-loops and duplicate edges. In our network construction, we fixed the rewiring probability at $p_{_{WS}}$ = 0.5.
The resulting network is denoted by $G(n,k_{_{WS}})$.

\item \textbf{Barab\'asi-Albert (BA) model} \cite{barabasi1999emergence}:
The BA model is used to generate scale-free networks characterized by a heterogeneous degree distribution. 
Starting from an initial network with $m_0$ vertices, where $m_0\geq m_{_{BA}}$, a new vertex is added at each step and connected to $m_{_{BA}}$ existing vertices. 
The vertices selected for attachment follow the preferential attachment rule, according to which vertices with larger degree have a higher probability of receiving new edges.
The resulting network is denoted by $G(n,m_{_{BA}})$.

\item \textbf{Hyperbolic Graph Generator (HGG) model}
\cite{krioukov2010hyperbolic}: The HGG model generates random hyperbolic networks that can simultaneously exhibit a power-law degree distribution and non-vanishing clustering. 
In this construction, $n$ vertices are distributed randomly in a hyperbolic disk, and edges are generated according to a probability that depends on the hyperbolic distance between pairs of vertices. 
The model is controlled by parameters such as the number of vertices $n$, the target average degree $k_{_{HGG}}$, the power-law exponent $\gamma$, and the temperature $T$.
In our analysis, the HGG networks are generated using the default parameter values $\gamma=2$ and $T=0$.
The resulting network is denoted by $G(n,k_{_{HGG}})$.

\end{itemize}

The structural characteristics of the model networks used in our analysis, including the number of vertices, number of edges, and fraction of vertices in
the largest connected component, are reported in Supplementary Table S1. For the computation and comparison of discrete curvatures, we restrict the analysis to the largest connected component of each network.

%=================================================================
\begin{table}[ht]
\centering
\caption{Real-world network characteristics, including the number of vertices and edges, the fraction of vertices in the largest connected component (LCC), and descriptions of vertices and edges for the networks considered in this study.}
\begin{tabular}{l c c c l l}
\toprule
\toprule
{\centering \textbf{Real-world Networks}} & \makecell[c]{\textbf{Number} \\ \textbf{of} \\ \textbf{Vertices}} & \makecell[c]{\textbf{Number} \\ \textbf{of} \\ \textbf{Edges}} & \makecell[c]{\textbf{Fraction of} \\\textbf{Vertices} \\ \textbf{in LCC}} & \textbf{Vertex description} & \textbf{Edge description}\\
\midrule
Adjective-Noun adjacency & \multirow{2}{*}{112} & \multirow{2}{*}{425} & \multirow{2}{*}{1} & \multirow{2}{*}{Nouns or adjectives} & Adjacency in the novel David \\
network \cite{newman2006finding} &  &  &  &  & Copperfield by Charles Dickens \\
\multirow{2}{*}{Chicago road network \cite{eash1979equilibrium}}& \multirow{2}{*}{1467}& \multirow{2}{*}{1298}& \multirow{2}{*}{0.561}& Transportation zones in & Road connections between \\
 &  &  &  & the Chicago region  & the zones\\
Contiguous US States & \multirow{2}{*}{49}& \multirow{2}{*}{107}& \multirow{2}{*}{1} & Contiguous states of & Land borders between two states\\
network \cite{knuth2006art} &  &  &  & the United States & \\
\multirow{2}{*}{Dolphin network \cite{lusseau2003bottlenose}}& \multirow{2}{*}{62}& \multirow{2}{*}{159}& \multirow{2}{*}{1} & Bottlenose dolphins in & Frequent associations between \\
 &  &  &  & Doubtful Sound, New Zealand  & the dolphins\\
Email communication & \multirow{2}{*}{1133}& \multirow{2}{*}{5451}& \multirow{2}{*}{1} & Users at the University Rovira & Direct communication between\\
 network \cite{guimera2003self} &  &  &  & i Virgili in Tarragona, Spain & users \\
\multirow{2}{*}{Euro road network \cite{vsubelj2011robust}}& \multirow{2}{*}{1174}& \multirow{2}{*}{1417}& \multirow{2}{*}{0.885}& \multirow{2}{*}{European cities}& Road connections in the \\
 &  &  &  &  & international E-road network\\
Facebook network \cite{leskovec2012learning} & 2888 & 2981 & 1 & Facebook users& Friendship relations among users\\
Football network \cite{girvan2002community} & 115 & 613 & 1 & American football teams& Games played between the teams\\
Hamsterster friendships & \multirow{2}{*}{1858}& \multirow{2}{*}{12534}& \multirow{2}{*}{0.962}& \multirow{2}{*}{Users of hamsterster.com}& \multirow{2}{*}{Friendship relations}\\
network &  &  &  &  & \\
Human protein interactions & \multirow{2}{*}{2783}& \multirow{2}{*}{6007}& \multirow{2}{*}{1} & \multirow{2}{*}{Proteins}& Documented interactions between\\ 
network \cite{rual2005towards} &  &  &  &  & human proteins\\
Jazz musicians network \cite{gleiser2003community} & 198 & 2742 & 1 & Jazz musicians& Collaborations between musicians\\
PDZ-domain interaction  & \multirow{2}{*}{212}& \multirow{2}{*}{242}& \multirow{2}{*}{0.759}& \multirow{2}{*}{Proteins}& PDZ-domain-mediated\\
network \cite{beuming2005pdzbase} &  &  &  &  & interactions between proteins\\
\multirow{2}{*}{Reactome network \cite{joshi2005reactome}} & \multirow{2}{*}{6327} & \multirow{2}{*}{147547} & \multirow{2}{*}{0.944} & \multirow{2}{*}{Human proteins} & Protein interactions obtained\\
 &  &  &  &  &  from the Reactome database\\
\multirow{2}{*}{Route views network \cite{leskovec2007graph}}& \multirow{2}{*}{6474}& \multirow{2}{*}{13895}& \multirow{2}{*}{1} & \multirow{2}{*}{Autonomous systems}& Communication links between \\
 &  &  &  &  & autonomous systems\\
\multirow{2}{*}{US Power Grid \cite{leskovec2007graph}}& \multirow{2}{*}{4941}& \multirow{2}{*}{6594}& \multirow{2}{*}{1} & Generators or transformers & \multirow{2}{*}{Power transmission lines}\\
 &  &  &  &  in the western United States & \\
Yeast protein interactions & \multirow{2}{*}{1846}& \multirow{2}{*}{2203}& \multirow{2}{*}{0.79}& Proteins of the yeast & Interactions between pairs\\
network \cite{jeong2001lethality}  &  &  &  & \textit{Saccharomyces cerevisiae} & of proteins\\
Zachary karate club & \multirow{2}{*}{34}& \multirow{2}{*}{78}& \multirow{2}{*}{1} & Members of a karate club& \multirow{2}{*}{Social ties between club members}\\
network \cite{zachary1977information}  &  &  &  &  studied by Wayne Zachary & \\
\multirow{2}{*}{Zebra network \cite{sundaresan2007network}}& \multirow{2}{*}{27}& \multirow{2}{*}{111}& \multirow{2}{*}{1} & \multirow{2}{*}{Grévy's zebras in Kenya}& Interactions between the animals\\
 &  &  &  &  & recorded during the study \\
\bottomrule
\end{tabular}
\label{Real_details}
\end{table}
%=================================================================

In addition to the model networks, we consider a collection of undirected and unweighted real-world networks from different application domains. 
The dataset includes infrastructure, biological, social, communication, linguistic, collaboration, online, computer, interaction, and animal networks. 
More specifically, the real-world dataset consists of four infrastructure networks, namely Chicago road \cite{eash1979equilibrium}, Contiguous US States \cite{knuth2006art}, Euro road \cite{vsubelj2011robust}, and US Power Grid \cite{leskovec2007graph}; four biological networks, namely Human protein interactions \cite{rual2005towards}, PDZ-domain \cite{beuming2005pdzbase}, Reactome \cite{joshi2005reactome}, and Yeast protein interactions \cite{jeong2001lethality}; four social networks, namely Facebook \cite{leskovec2012learning}, Hamsterster friendships, Jazz musicians \cite{gleiser2003community}, and Zachary karate club \cite{zachary1977information}. 
We further include an Adjective-Noun adjacency network \cite{newman2006finding}, an Email communication network \cite{guimera2003self}, a Football interaction network \cite{girvan2002community}, a Route views computer network \cite{leskovec2007graph}, and two animal social networks, Dolphin \cite{lusseau2003bottlenose} and Zebra \cite{sundaresan2007network}.
Most of the undirected real-world networks used in this study were obtained from the KONECT database \cite{kunegis2013konect}. 
The main characteristics of these networks are summarised in Table \ref{Real_details}. 
Since all networks considered here are unweighted, we assign unit weights to vertices, edges, and two-dimensional simplicial complexes when computing the different curvature quantities.

%=============================================================

%=======================================================
\section{Results and Discussion}\label{sec:results}
%============================================================

In this study, we explored Steinerberger curvature (SBC), as well as a notion of resistance curvature (RC), across different network structures and examined its relationship with established discrete Ricci curvature measures, namely Forman-Ricci curvature (FRC), Augmented Forman-Ricci curvature (AFRC), and Ollivier-Ricci curvature (ORC), along with scalar sectional curvature (SSC) and with vertex-based graph measures. 
To this end, we considered four synthetic network models with systematically varied parameters, along with 18 undirected and unweighted real-world networks.
For each network model, the Steinerberger curvature was evaluated across three network sizes ($n$ = 1000, 2000, and 5000), across systematically varied model-specific parameters to capture differences in network architecture.
The parameter settings were as follows: $p_{_{ER}} \in \{3/n,\ 5/n,\ 7/n,\ 10/n\}$ for the ER model $G(n,p_{_{ER}})$; $k_{_{WS}} \in$ \{2, 4, 6, 8, 10\} for the WS model $G(n,k_{_{WS}})$; $m_{_{BA}} \in$ \{2, 3, 4, 5\} for the BA model $G(n,m_{_{BA}})$; and $k_{_{HGG}} \in$ \{3, 5, 7, 9, 10\} for the HGG model  $G(n,k_{_{HGG}})$.
Corresponding to each combination of network size and model-specific parameter, 100 independent random networks were generated. 
For all analyses, we restricted the networks to their largest connected component.
This systematic sampling allowed us to examine the distribution and variation of SBC and RC across different network topologies, and subsequently assess their correspondence with discrete Ricci curvatures and other vertex-based graph measures.

%=============================================================
\subsection{Basic characteristics of Steinerberger curvature in model and real-world networks}

%=============================================================
% Figure 1
\begin{figure}
    \centering
    \includegraphics{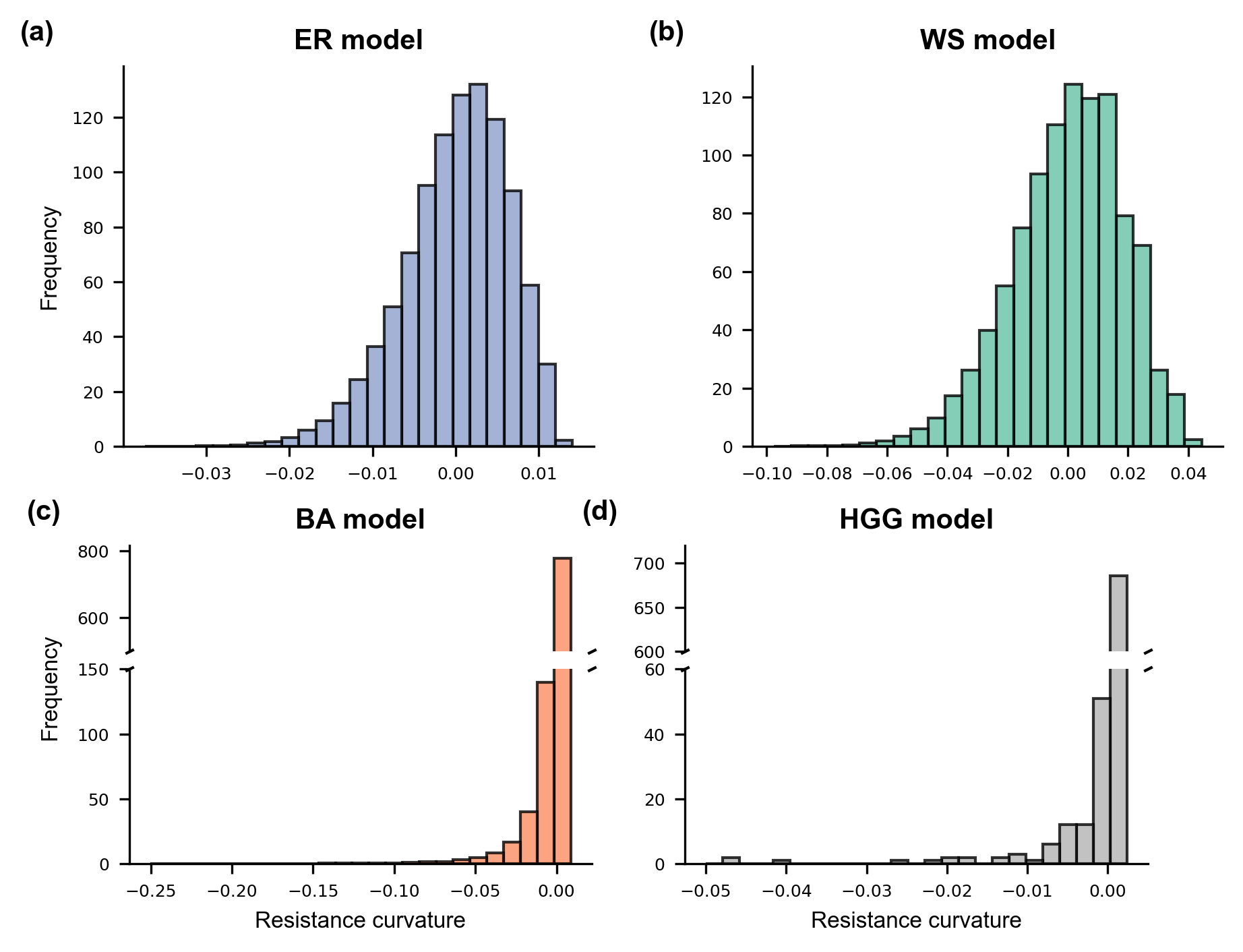}
    \caption{Distribution of resistance curvatures across 100 independently generated model networks for each parameter combination, with $n=1000$ vertices: (a) ER model with $p_{\mathrm{ER}}=0.005$, (b) WS model with $k_{\mathrm{WS}}=6$, (c) BA model with $m_{\mathrm{BA}}=3$, and (d) HGG model with $k_{\mathrm{HGG}}=3$. A break in the y-axis is applied to accommodate the wide range of the frequency distributions.}
    \label{fig:ESBC_distribution_Model}
\end{figure}
%=============================================================

We first examined the distributions of SBC and RC across different model networks and real-world networks to understand their variation across different network structures.  
For the ER, WS, and BA models, the SBC values generally followed a bell-shaped distribution centered near zero, indicating that most nodes had SBC values close to zero. 
However, the distribution of SBC varied considerably across network models and independently generated networks with the same model and parameter combination, ranging from relatively narrow to substantially broader distributions.
This variation suggested that although the overall distributional shape remained similar, the magnitude of SBC could be strongly influenced by the specific network architecture. 
In contrast, the HGG model exhibited a markedly different distribution, with SBC highly concentrated around zero and only a small number of observations showing substantially negative or positive values.  
Supplementary Figure S1 illustrates the mean distribution of SBC for the four model networks with $n$ = 1000 nodes, considering one representative parameter setting for each model.

Figure \ref{fig:ESBC_distribution_Model} illustrates the mean distributions of RC across four model networks, each consisting of $n$ = 1000 vertices. 
In the ER and WS models, RC values followed a bell-shaped distribution centered near zero with a slight extension towards negative values, indicating that most nodes had RC values close to zero.  
In contrast, for BA and HGG models, RC values followed a right-skewed distribution.
The BA model displayed a highly concentrated distribution near zero, together with a pronounced left tail extending towards significantly negative values. 
This indicates that while most vertices possess curvature values close to zero, a relatively small proportion exhibit substantially negative curvature. 
In the HGG model, the curvature values are strongly concentrated near zero, accompanied by a relatively shorter negative tail. 
The broken vertical axes in the BA and HGG histograms highlight the substantial concentration of observations near zero compared with the frequencies in the tails. 
Overall, these results indicated that both SBC and RC values were generally concentrated around zero, although their spread and the occurrence of extreme values varied across network models. 
Furthermore, SBC exhibits a wider distribution, whereas RC is more narrowly concentrated.
These observations suggest that network structure influences the distribution of both curvature measures, with differences in their distributional patterns across the considered models.

The distributions of SBC and RC for the different real-world networks are shown in Supplementary Figures S2 and S3, respectively. 
Most networks showed SBC values concentrated around zero with approximately bell-shaped distributions, while RC values generally clustered around zero and exhibited right-skewed distributions. 
These patterns were broadly consistent with those observed in the model networks, although the distributions varied across real-world networks.

%=====================================================================

%=====================================================================
\subsection{Correlation between Steinerberger curvature and other measures}

In this section, we examine the relationship of SBC and RC with established discrete Ricci curvature measures, SSC, and conventional vertex-based measures across model and real-world networks.
In any model network, for each parameter setting, we generated 100 network realizations and computed the Pearson and Spearman correlations between SBC and each vertex-based graph measure for every network. 
Then the mean correlation across all 100 networks was calculated, restricting all analyses to the largest connected component (LCC) of each network.

%=====================================================================
\subsubsection{Comparison with discrete Ricci curvatures}

% Figure 2
\begin{figure}
    \centering
    \includegraphics[width=0.95\linewidth]{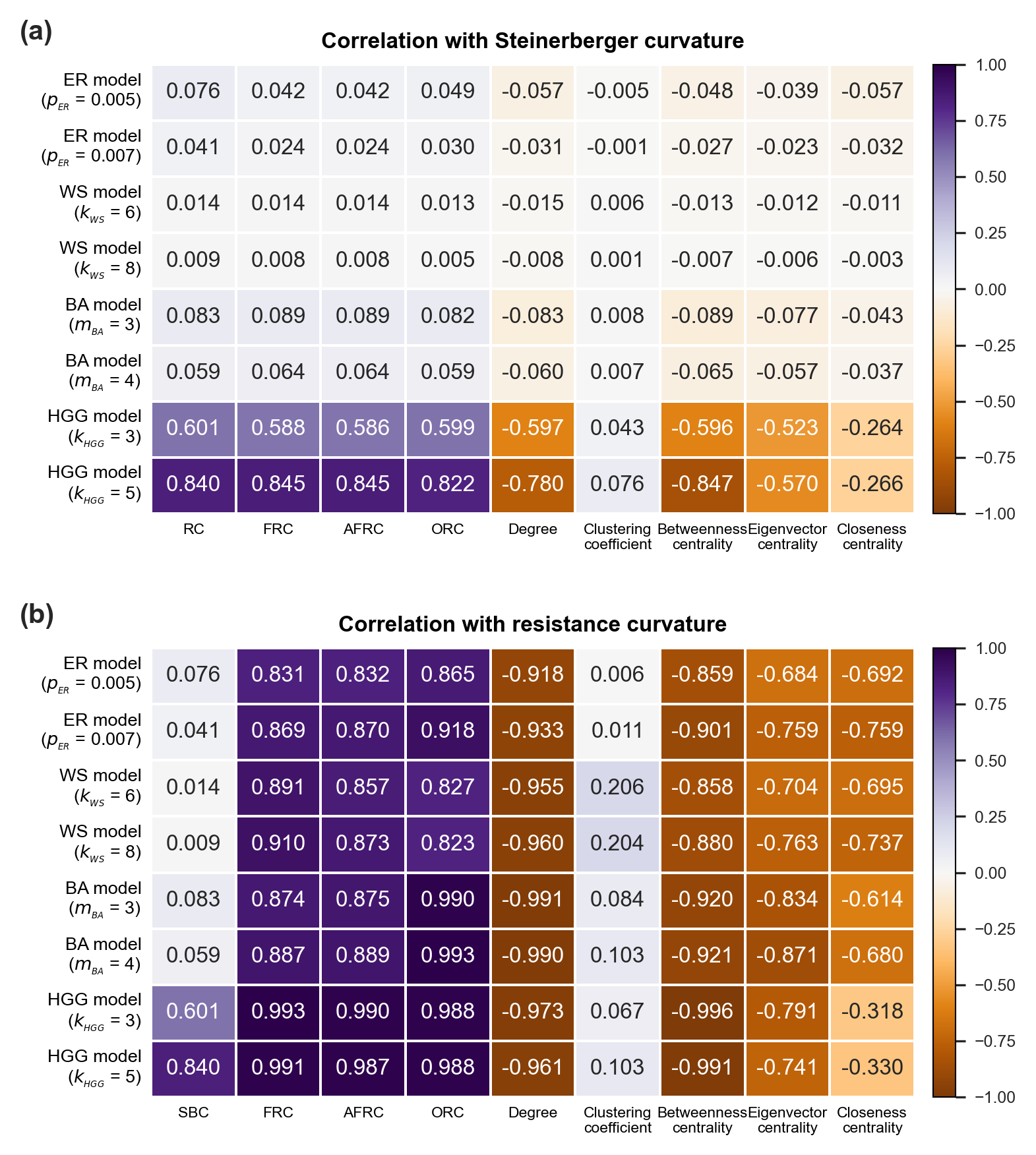}
    \caption{Heatmaps illustrating Pearson correlation (rounded to three decimal places) of (a) Steinerberger curvature (SBC) and (b) resistance curvature (RC) with Forman-Ricci curvature (FRC), Augmented Forman-Ricci curvature (AFRC), Ollivier-Ricci curvature (ORC), degree, clustering coefficient, betweenness centrality, eigenvector centrality, and closeness centrality in model networks. Each column represents a different vertex measure, and each row corresponds to a specific network model and parameter combination.}
    \label{fig:Correlation_Model}
\end{figure}

% Figure 3
\begin{figure}
    \centering
    \includegraphics[width=0.95\linewidth]{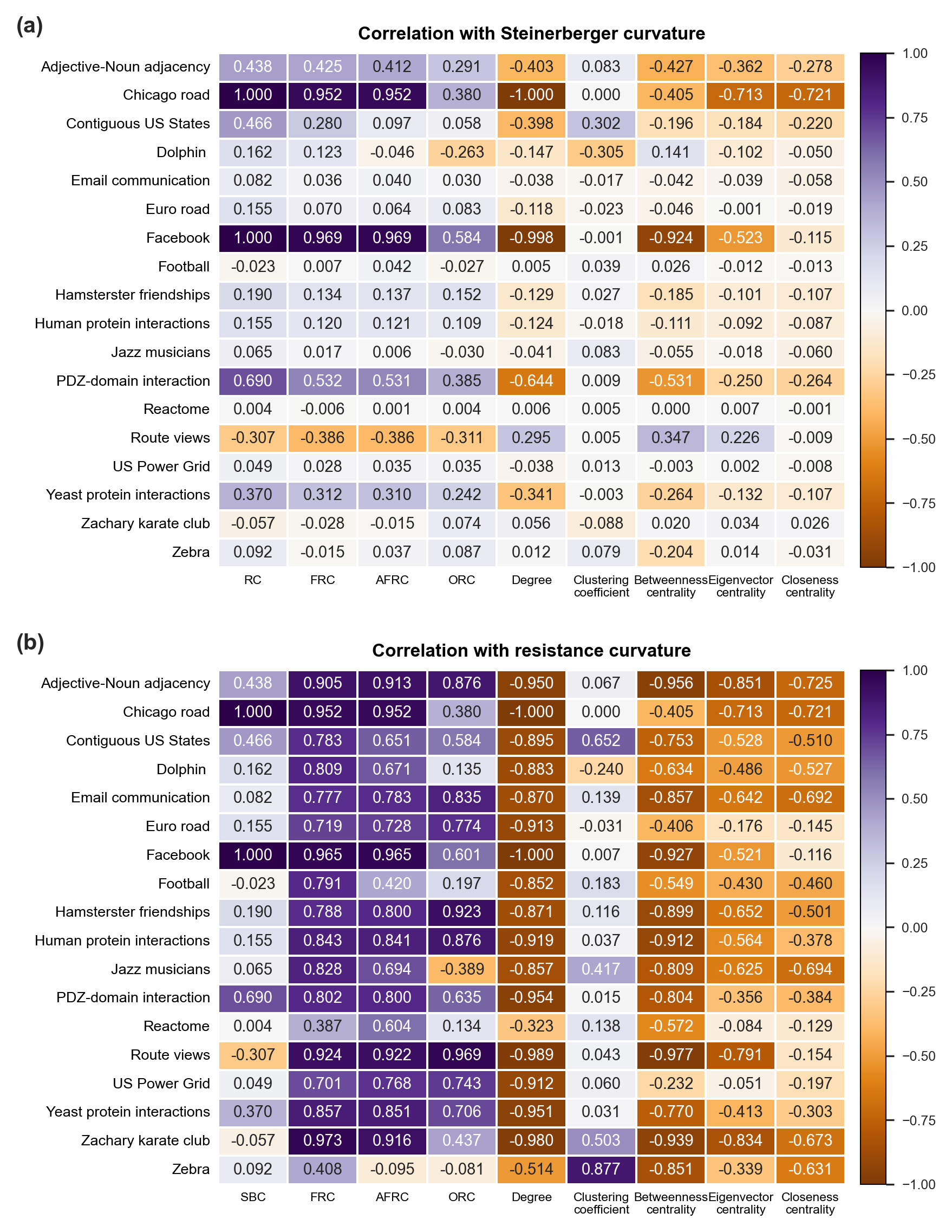}
    \caption{Heatmaps illustrating Pearson correlation (rounded to three decimal places) of Steinerberger curvature (SBC) and (b) resistance curvature (RC) with Forman-Ricci curvature (FRC), Augmented Forman-Ricci curvature (AFRC), Ollivier-Ricci curvature (ORC), degree, clustering coefficient, betweenness centrality, eigenvector centrality, and closeness centrality in model networks. The columns correspond to the nine different vertex measures. The rows correspond to different real-world networks.}
    \label{fig:Correlation_Real}
\end{figure}

Next, we investigated the relationship between SBC and established discrete Ricci curvature measures, namely, FRC, AFRC, and ORC. 
For the ER, WS, and BA models, the correlations were generally weak and close to zero across all three curvature measures, suggesting that SBC captured structural information that was largely distinct from that represented by these established measures. 
In contrast, the HGG model exhibited moderate to strong positive correlations, with coefficients ranging approximately from 0.586 to 0.845 (see Figure \ref{fig:Correlation_Model}). 
Furthermore, FRC and AFRC yielded identical or very similar correlation values in most cases, whereas ORC showed some differences in magnitude. 
The stronger correlations observed in the HGG model may be related to its comparatively shorter average shortest path length, which could influence the relationship between distance-based and Ricci curvature measures. 
However, further analysis is required to establish whether this network property explains the observed correlations. 
Overall, these findings suggest that the relationship between SBC and discrete Ricci curvature-based measures depends on network structure, with limited associations in most of the considered models and a notably stronger relationship in the HGG model.
The second, third, and fourth columns of the heatmap in Figure \ref{fig:Correlation_Model}(a) show the Pearson correlations between SBC and the three discrete Ricci curvature-based measures. 
For all model networks analyzed, both Pearson and Spearman correlation coefficients between SBC and discrete Ricci curvature-based measures are reported in Supplementary Tables S2 and S3.

For the real-world networks, the correlations between SBC and the established discrete Ricci curvature measures were generally weak or close to zero, indicating limited associations between these measures across most networks.
However, the strength of the correlations varied across individual networks.  
These findings indicate that the relationship between curvature measures may depend on the structural characteristics of the underlying real-world networks. 
The second, third and fourth columns of the heatmap in Figure \ref{fig:Correlation_Real}(a) present the Pearson correlation coefficients for the three curvature measures, while the corresponding Pearson and Spearman correlation coefficients for all real-world networks are reported in Supplementary Table S4.

The first four columns of the heatmap in Figure \ref{fig:Correlation_Model}(b) illustrate the Pearson correlation between the RC and other curvature-based measures across eight synthetic networks generated using four standard random graph models.
Remarkably, RC exhibits strong positive correlations ($>0.8$) with all three Ricci curvature-based measures across all model networks, most prominently with ORC.
In contrast, the relationship with standard SBC varies considerably across network models. 
While ER, WS, and BA networks show very weak correlations, HGG networks exhibit moderate to strong positive correlations. 
This difference suggests that the underlying hyperbolic or metric geometry fundamentally shapes the relationship between effective resistance and shortest path distance-based Steinerberger curvature.
Both Pearson and Spearman correlation coefficients for all model networks are reported in Supplementary Tables S5 and S6.

Similar to the model networks, most real-world networks exhibit strong positive correlations between RC and Ricci curvature-based measures.
The correlations are more pronounced for FRC. 
In contrast, the correlations with ORC vary across networks, ranging from weak to strong. 
The relationship between RC and SBC is generally weakly positive, although in some networks moderate to high correlations are observed.
The first four columns of the heatmap in Figure \ref{fig:Correlation_Real}(b) show the Pearson correlation coefficients of RC with the other four curvature measures.
Corresponding Pearson and Spearman correlation coefficients for all real-world networks are reported in Supplementary Table S7.

%=====================================================================

%=====================================================================
\subsubsection{Comparison with scalar sectional curvature}

Next, we compared SBC and RC with SSC across both model and real-world networks. 
Given the high computational cost of evaluating SSC, for the model networks, we restricted the correlation analysis to networks with $n$ = 1000 vertices and calculated the correlations for 10 random networks for each model-specific parameter setting.
We extended the analysis to 12 real-world networks to examine the relationship between the curvature measures.
The results highlight that the correlations with SBC were generally close to zero in most cases, indicating weak associations between the two curvature measures. 
In contrast, SSC generally exhibits a weak to moderate negative correlation with RC across both model and real-world networks. 
These findings suggest that SSC has a different relationship with RC than with SBC. 
Although all three measures provide geometric perspectives on network structure, their correlations indicate that they capture different structural characteristics in most of the networks considered in this study.
For both the model and real-world networks, the Pearson and Spearman correlation coefficients of SSC with SBC and RC are reported in Supplementary Tables S8 and S9, respectively.

%=====================================================================

%=====================================================================
\subsubsection{Comparison with vertex-based graph measures}

We examined the relationship between SBC and five commonly used vertex-based graph measures: degree, clustering coefficient, betweenness centrality, eigenvector centrality \cite{bonacich2007some}, and closeness centrality \cite{okamoto2008ranking}. 
Degree captured the number of connections of a vertex to its immediate neighbors and reflected its local connectivity.
Clustering coefficient described the extent to which a vertex’s neighbors were also connected to one another. 
Betweenness centrality quantified the role of a vertex in connecting other parts of the network by measuring its occurrence on shortest paths. 
Eigenvector centrality reflected vertex importance based on both its connections and the importance of its neighboring vertices, whereas closeness centrality captured the proximity of a vertex to the rest of the network based on shortest path distances.
Overall, these measures capture different aspects of the connectivity, local structure, and topological position of individual vertices within a network.

The correlations between SBC and the considered graph measures were generally close to zero for the ER, WS, and BA models, with most correlation coefficients ranging approximately from -0.01 to -0.1 (see the last five columns in Figure \ref{fig:Correlation_Model}(a)). 
In contrast, the HGG model exhibited a markedly different pattern. 
In this model, SBC showed moderate to strong negative correlations with degree, betweenness centrality, and eigenvector centrality, whereas its correlations with clustering coefficient and closeness centrality were relatively weak and negative. 
Thus, while SBC showed only weak associations with the conventional graph measures in the ER, WS, and BA models, its relationship with these measures was more pronounced in the HGG model. 
These results indicated that the association between SBC and conventional vertex-based measures may depend on the underlying network structure.
In particular, the stronger correlations observed in HGG suggested that SBC was more closely related to connectivity and centrality in this network model with explicit geometric structure. 
In contrast, it remained largely distinct from the considered graph measures in the ER, WS, and BA networks.
The mean and standard deviation of both the Pearson and Spearman correlation coefficients for all the model networks are provided in Supplementary Tables S10–S12.

For the real-world networks, the correlation patterns were broadly consistent with those observed for the synthetic networks, with most correlations remaining relatively weak (see the last five columns in Figure \ref{fig:Correlation_Real}(a)). 
However, several networks showed substantially stronger associations for degree- and centrality-based measures. 
Overall, these results indicated that although SBC was largely distinct from the considered vertex-based measures in most real-world networks, stronger relationships emerged in a subset of networks, particularly for degree, betweenness centrality, and eigenvector centrality.
The Pearson and Spearman correlation coefficients for all 18 real-world networks are provided in Supplementary Table 13.

On the other hand, the relationship between RC and conventional graph measures presents a different pattern.
Among these, degree and betweenness centrality exhibit consistently strong negative correlations ($<-0.85$) with RC across all network models. 
Eigenvector and closeness centralities also show moderate to strong negative correlations. 
In contrast, the clustering coefficient exhibits either no correlation or only weak positive correlations with RC.
The last five columns of the heatmap in Figure \ref{fig:Correlation_Model}(b) illustrate the Pearson correlations between RC and five conventional graph measures in the model networks. 
A similar pattern is observed in the real-world networks as well (see the last five columns of Figure \ref{fig:Correlation_Real}(b)). 
However, the negative correlation with degree is more consistent than that with betweenness centrality across these networks. 
Overall, these results demonstrate that the strong negative association with degree and betweenness centrality, remains remarkably consistent across different random graph models and most of the real-world networks.
The corresponding mean and standard deviation for the Pearson and Spearman correlations are reported in Supplementary Tables S14–S16 for the model networks and in Supplementary Table S17 for the real-world networks.

%=====================================================================
\subsection{The relative importance of Steinerberger curvature and topological robustness of networks}

% Figure 4
\begin{figure}
    \centering
    \includegraphics{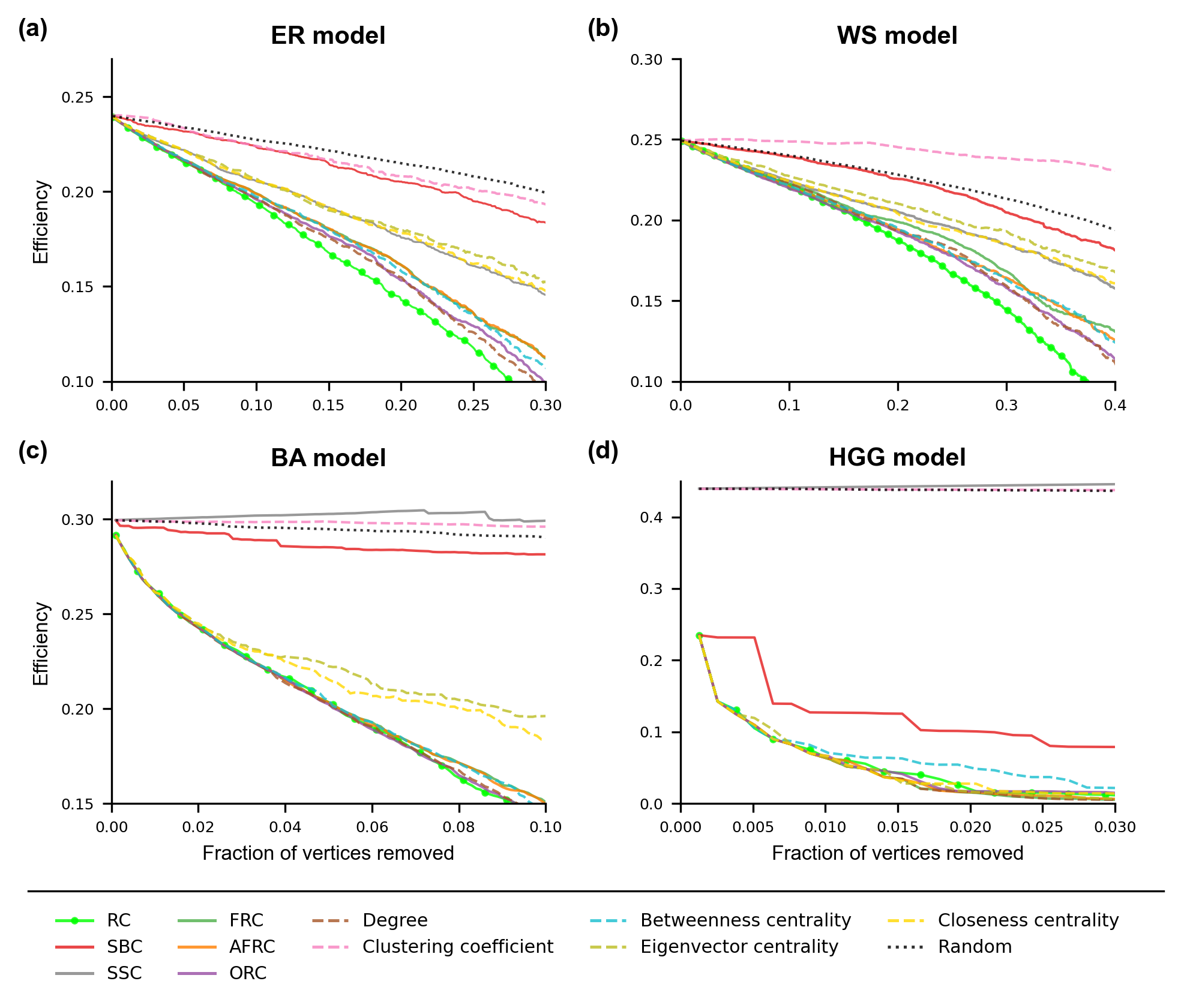}
    \caption{Communication efficiency versus the fraction of vertices removed in model networks with $n$ = 1000 vertices. (a) ER model with $p_{_{ER}}$ = 0.005, (b) WS model with $k_{_{WS}}$ = 6, (c) BA model with $m_{_{BA}}$ = 3, and (d) HGG model with $k_{_{HGG}}$ = 3. Solid, dashed, and dotted lines correspond to different curvature measures, vertex measures, and random vertex removal strategies, respectively. Abbreviations: RC: Resistance curvature, SBC: Steinerberger curvature, SSC: scalar sectional curvature, FRC: Forman-Ricci curvature, AFRC: Augmented Forman-Ricci curvature, ORC: Ollivier-Ricci curvature.}
    \label{fig:Robustness_Model}
\end{figure}

% Figure 5
\begin{figure}
    \centering
    \includegraphics[width=1\linewidth]{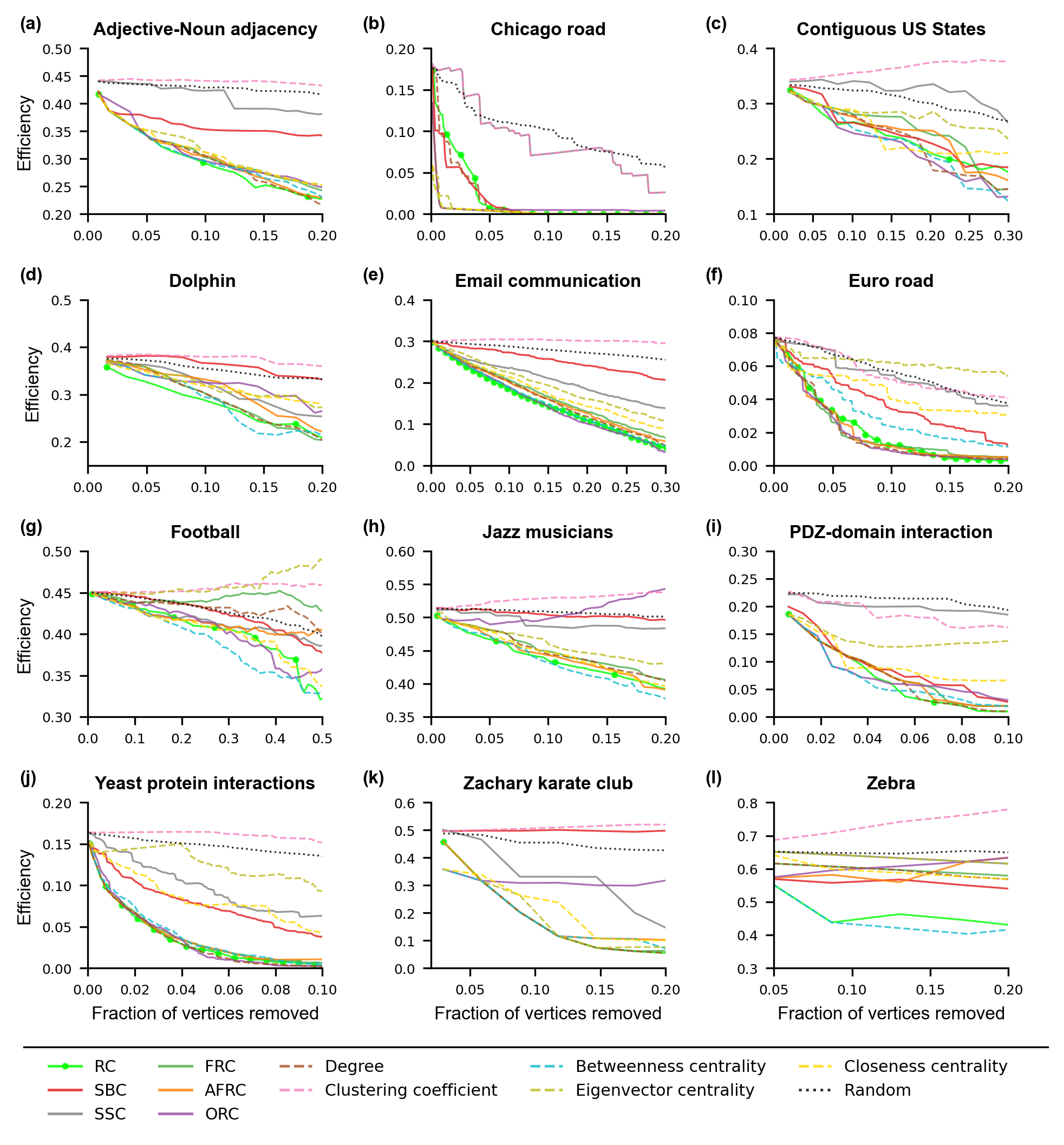}
    \caption{Communication efficiency versus the fraction of vertices removed in real-world networks. (a) Adjective-Noun Adjacency, (b) Chicago Road, (c) Contiguous US States, (d) Dolphin, (e) Email Communication, (f) Euro Road, (g) Football, (h) Jazz Musicians, (i) PDZ-Domain Interaction, (j) Yeast Protein Interactions, (k) Zachary Karate Club, and (l) Zebra networks. Solid, dashed, and dotted lines correspond to different curvature measures, vertex measures, and random vertex removal strategies, respectively. Abbreviations: RC: Resistance curvature, SBC: Steinerberger curvature, SSC: scalar sectional curvature, FRC: Forman-Ricci curvature, AFRC: Augmented Forman-Ricci curvature, ORC: Ollivier-Ricci curvature.}
    \label{fig:Robustness_Real}
\end{figure}

Communication efficiency is a global network measure that quantifies how efficiently information can be exchanged between pairs of vertices through the shortest paths. 
It is defined as the average of the inverse shortest path distances between all distinct pairs of vertices in a network. For an undirected network $G=(V,E)$ with $n$ vertices, communication efficiency is expressed as
\begin{equation*}
 E(G) = \frac{1}{n(n-1)}\sum_{i \neq j \in V(G)}{\frac{1}{d(i,j)}},
\end{equation*}
where, $d(i,j)$ denotes the shortest path distance between vertices $i$ and $j$. 
If no path exists between a pair of vertices, their contribution is defined as zero, with $1/d(i,j)=0$ when $d(i,j)=\infty$. 
Higher communication efficiency indicates that vertices can reach one another through shorter paths, whereas lower efficiency suggests less efficient communication across the network. 

Finally, we investigated how network communication efficiency changes as vertices are progressively removed. 
We considered 12 vertex measures, along with random vertex removal, and calculated the communication efficiency at each step of the removal process. 
For RC, SBC, FRC, AFRC, and ORC, vertices were removed in increasing order of curvature values, without recalculating the values after each removal. 
In contrast, for degree, clustering coefficient, betweenness centrality, eigenvector centrality, and closeness centrality, vertices were removed in decreasing order of the corresponding measure.
SSC was also considered, with vertices removed in decreasing order, as it exhibited negative correlations with curvature-based measures and positive correlations with graph-based measures.
For random removal, vertices were removed one at a time in a randomly selected order. 
This procedure was repeated ten times for each network, and the mean communication efficiency at each removal step was calculated across the ten runs.

The results for the model networks show that RC achieves the fastest network disintegration in the ER and WS models, outperforming all other measures.
In BA and HGG networks, RC follows a trend similar to that of the established Ricci curvature-based measures.
Furthermore, Ricci curvature-based measures consistently outperform SBC and SSC in accelerating network disintegration (see Figure \ref{fig:Robustness_Model}). 
Consistent with previous studies \cite{samal2018comparative}, ORC, FRC, and AFRC exhibit broadly similar performance to degree and betweenness centrality. 
However, the relative performance of SBC and SSC varies across network types.
SBC consistently outperforms random removal and clustering coefficient across all models. 
SSC performs better than SBC in ER and WS networks.
However, it shows performance comparable to random removal and clustering coefficient in BA and HGG networks.
Eigenvector centrality and closeness centrality also exhibit similar performance patterns, with intermediate effectiveness among the considered strategies.
Overall, the results suggest that vertices with more negative RC values play an important role in maintaining network connectivity. 
RC performs particularly well in ER and WS networks, while its performance in BA and HGG networks follows a pattern similar to the established Ricci curvature measures.
The consistently better performance of RC compared to SBC suggests that effective resistance distance captures network structure more effectively than shortest path distance in identifying important vertices.
Figure \ref{fig:Robustness_Model} illustrates the communication efficiency of all four model networks under different vertex removal strategies.

We performed this analysis for real-world networks as well. 
The results reveal that the patterns are broadly consistent with those observed in model networks. 
Figure \ref{fig:Robustness_Real} illustrates the changes in communication efficiency under different vertex removal strategies for the real-world networks.
In most cases, RC along with discrete Ricci curvature-based measures, degree and betweenness centrality, generally lead to faster declines in communication efficiency compared to SBC and SSC.
However, SBC and SSC outperform random vertex removal and clustering coefficient, indicating their greater effectiveness in accelerating network disintegration compared with these strategies. 
Although the relative performance of individual measures varies across networks, these findings highlight the effectiveness of curvature-based approaches and conventional centrality measures in identifying influential vertices and accelerating the disintegration of real-world networks.

%==============================================================

%==============================================================
\section{Conclusion}\label{sec:conclusion}

Curvature provides a useful mathematical framework for understanding the structural properties of graphs and complex networks. Among the distance-based approaches, SBC, introduced through the equilibrium measure associated with the graph distance matrix, provides a global characterization of vertices by incorporating pairwise shortest path distances. 
Building on this framework, another notion of resistance curvature is introduced, which is essentially equivalent to the existing resistance curvature \cite{devriendt2022discrete}, by replacing the graph distance matrix with the effective resistance matrix.
Unlike shortest path distance, effective resistance distance accounts for the contributions of all available paths between pairs of vertices and therefore incorporates information about the global connectivity of the network. 
Motivated by this distinction, in this work, we investigated SBC and its extension, based on RC. 
The distributional analysis indicated that both SBC and RC values were generally concentrated around zero, with variations depending on the underlying network model. 
Furthermore, we examined their relationships with established measures, including FRC, AFRC, ORC, and SSC, as well as conventional vertex-based graph measures, such as degree, clustering coefficient, betweenness centrality, eigenvector centrality, and closeness centrality, across real-world and model networks. 

Notably, the computational time differs across the considered curvature-based measures, which has implications for their application to large networks.
Both SBC and RC involve dense matrix operations that make their computation relatively expensive. 
In comparison, FRC and AFRC are relatively inexpensive to compute since they depend on local graph properties. 
On the other hand, ORC requires solving an optimal transport problem for each edge, which makes it more computationally demanding.
SSC is also computationally demanding, as it requires examining all vertex triplets across possible side lengths, followed by iterative ball expansion until they share a common intersection.
Typically, for a given network, FRC has the lowest computational cost, followed by AFRC, SBC, RC, and ORC, while SSC is generally the most computationally expensive. 

Our comparative analysis revealed strong positive correlations between RC and discrete Ricci curvatures. 
In contrast, RC exhibited predominantly negative correlations with vertex-based graph measures, whereas the correlations of SBC with established discrete Ricci curvature measures and vertex-based graph measures were generally weak, indicating limited associations across most networks.
Furthermore, the communication efficiency study revealed that SBC consistently outperformed random removal and the clustering coefficient across all considered models, although it did not outperform other curvature-based measures. 
In comparison, RC generally showed strong performance, either outperforming or performing comparably to established discrete Ricci curvature measures (ORC, FRC, and AFRC), as well as degree and betweenness centrality.
It also consistently outperformed the remaining vertex-based measures considered in this study.

Overall, this study highlights the importance of the underlying distance measure in defining distance-based graph curvature. 
The observed differences between SBC and RC suggest that effective resistance, through its incorporation of information from multiple paths, provides a potentially richer perspective on network connectivity than shortest path distance alone. 
These findings motivate further theoretical investigations into the relationship between effective resistance-based curvature and other notions of discrete curvature, as well as its applications in characterizing and analysing complex network structures.

\section*{COMPETING INTERESTS}
The authors declare no conflicts of interest.

\section*{AUTHOR CONTRIBUTIONS}
Designed the research: M.M., P.V., A.S., J.J.; Performed the research: M.M., P.V., A.S., J.J.; Performed the computations: M.M.; Wrote the paper: M.M., P.V., A.S., J.J.

\section*{ACKNOWLEDGMENTS}
The authors thank Yasharth Yadav for insightful discussions. A.S. would like to acknowledge funding from the Department of Atomic Energy (DAE), Government of India via Advanced Research and Education in Mathematical Sciences at IMSc [O.M. No. 2103/1/2025-R\&D-II-DAE/IMSc/4885 dated 23.4.2026].

%==============================================================

%==============================================================
% References
\bibliography{refs}
\bibliographystyle{ieeetr}
%==============================================================

%==============================================================
\newpage
\onecolumngrid
\setcounter{page}{1}
\setcounter{table}{0}
\renewcommand{\thetable}{S\arabic{table}}
\setcounter{figure}{0}
\renewcommand{\thefigure}{S\arabic{figure}}
\setcounter{section}{0}
\renewcommand{\thesection}{\arabic{section}}

\setcounter{secnumdepth}{1}

\title{SUPPLEMENTARY INFORMATION\\ for \\ A comparative exploration of Steinerberger curvature and resistance curvature in model and real-world complex networks}
\maketitle

%==================================================
%SUPPLEMENTARY TABLES
%==============================================
%==================================================
\section{SUPPLEMENTARY TABLES}

% Table S1
\begin{table}
\centering
\caption {Model network characteristics, including the number of vertices and edges, the fraction of vertices in the largest connected component (LCC), and the average shortest path length for the networks considered in this study.}
\begin{tabular}{l c c c c}

\toprule
\toprule
{\centering \textbf{Model Networks}} & \makecell[c]{\textbf{Number} \\ \textbf{of vertices}} & \makecell[c]{\textbf{Number} \\ \textbf{of edges}} & \makecell[c]{\textbf{Fraction of vertices} \\ \textbf{in LCC}}  & \makecell[c]{\textbf{Average shortest} \\ {\textbf{path length}}} \\
\midrule
\textbf{ER ($n$ = 1000, $p_{_{ER}}$ = 0.003)} & 1000 & 1503 & 0.942&6.228\\
\textbf{ER ($n$ = 1000, $p_{_{ER}}$ = 0.005)} & 1000 & 2506 & 0.993&4.457\\
\textbf{ER ($n$ = 1000, $p_{_{ER}}$ = 0.007)} & 1000 & 3504 & 0.999&3.765\\
\textbf{ER ($n$ = 1000, $p_{_{ER}}$ = 0.01)} & 1000 & 4992 & 1&3.258\\
\textbf{ER ($n$ = 2000, $p_{_{ER}}$ = 0.0015)} & 2000 & 3004 & 0.942&6.871\\
\textbf{ER ($n$ = 2000, $p_{_{ER}}$ = 0.0025)} & 2000 & 5006 & 0.993&4.892\\
\textbf{ER ($n$ = 2000, $p_{_{ER}}$ = 0.0035)} & 2000 & 7002 & 0.999&4.126\\
\bottomrule
\end{tabular}
\label{Model_details}
\end{table}

% Table S2
\begin{sidewaystable}
\centering
\caption{Pearson and Spearman correlation along with standard deviation of Steinerberger curvature with Forman-Ricci, Augmented Forman-Ricci, and Ollivier-Ricci curvature of vertices in the ER and WS model networks generated with different parameter values. The correlations are given up to three decimal places.}
\begin{tabular}{l@{\hspace{0.5cm}} c@{\hspace{0.4cm}} c@{\hspace{0.5cm}} c@{\hspace{0.4cm}} c@{\hspace{0.5cm}} c@{\hspace{0.4cm}} c@{\hspace{0.5cm}}}
\toprule
\toprule
\multirow{2}{*}{\textbf{Model Networks}} & \multicolumn{2}{c}{\textbf{Forman-Ricci}} & \multicolumn{2}{c}{\textbf{Augmented Forman-Ricci}} & \multicolumn{2}{c}{\textbf{Ollivier-Ricci}} \\
\cmidrule(lr){2-3}
\cmidrule(lr){4-5}
\cmidrule(lr){6-7}
& \textbf{~~Pearson~~} & \textbf{~~Spearman~~~} & \textbf{~~Pearson~~} & \textbf{~~Spearman~~} & \textbf{~~Pearson~~} & \textbf{~~Spearman~~~} \\
\midrule
\textbf{ER ($n$ = 1000, $p_{_{ER}}$ = 0.003)} & 0.077 $\pm$ 0.063 & 0.122 $\pm$ 0.078 & 0.077 $\pm$ 0.063 & 0.122 $\pm$ 0.078 & 0.084 $\pm$ 0.063 & 0.116 $\pm$ 0.076 \\
\textbf{ER ($n$ = 1000, $p_{_{ER}}$ = 0.005)} & 0.042 $\pm$ 0.039 & 0.054 $\pm$ 0.041 & 0.042 $\pm$ 0.039 & 0.054 $\pm$ 0.041 & 0.049 $\pm$ 0.042 & 0.051 $\pm$ 0.040 \\
\textbf{ER ($n$ = 1000, $p_{_{ER}}$ = 0.007)} & 0.024 $\pm$ 0.030 & 0.025 $\pm$ 0.030 & 0.024 $\pm$ 0.030 & 0.025 $\pm$ 0.030 & 0.030 $\pm$ 0.032 & 0.026 $\pm$ 0.032 \\
\textbf{ER ($n$ = 1000, $p_{_{ER}}$ = 0.01)} & 0.023 $\pm$ 0.032 & 0.023 $\pm$ 0.032 & 0.023 $\pm$ 0.032 & 0.023 $\pm$ 0.032 & 0.025 $\pm$ 0.035 & 0.023 $\pm$ 0.033 \\
\textbf{ER ($n$ = 2000, $p_{_{ER}}$ = 0.0015)} & 0.084 $\pm$ 0.054 & 0.128 $\pm$ 0.074 & 0.084 $\pm$ 0.054 & 0.128 $\pm$ 0.074 & 0.092 $\pm$ 0.056 & 0.123 $\pm$ 0.070 \\
\textbf{ER ($n$ = 2000, $p_{_{ER}}$ = 0.0025)} & 0.033 $\pm$ 0.030 & 0.047 $\pm$ 0.031 & 0.033 $\pm$ 0.030 & 0.047 $\pm$ 0.032 & 0.039 $\pm$ 0.032 & 0.044 $\pm$ 0.031 \\
\bottomrule
\end{tabular}
\end{sidewaystable}

% Table S3
\begin{sidewaystable}
\centering
\caption{Pearson and Spearman correlation along with standard deviation of Steinerberger curvature with Forman-Ricci, Augmented Forman-Ricci, and Ollivier-Ricci curvature of vertices in the BA and HGG model networks generated with different parameter values. The correlations are given up to three decimal places.}
\begin{tabular}{l@{\hspace{0.5cm}} c@{\hspace{0.4cm}} c@{\hspace{0.5cm}} c@{\hspace{0.4cm}} c@{\hspace{0.5cm}} c@{\hspace{0.4cm}} c@{\hspace{0.5cm}}}
\toprule
\toprule
\multirow{2}{*}{\textbf{Model Networks}} & \multicolumn{2}{c}{\textbf{Forman-Ricci}} & \multicolumn{2}{c}{\textbf{Augmented Forman-Ricci}} & \multicolumn{2}{c}{\textbf{Ollivier-Ricci}} \\
\cmidrule(lr){2-3}
\cmidrule(lr){4-5}
\cmidrule(lr){6-7}
& \textbf{Pearson~~~~} & \textbf{Spearman~~~~~} & \textbf{Pearson~~~~} & \textbf{Spearman~~~~~} & \textbf{Pearson~~~~} & \textbf{Spearman~~~~~} \\
\midrule
\textbf{BA ($n$ = 1000, $m_{_{BA}}$ = 2)} & 0.051 $\pm$ 0.172 & 0.015 $\pm$ 0.034 & 0.051 $\pm$ 0.172 & 0.015 $\pm$ 0.034 & 0.058 $\pm$ 0.126 & 0.029 $\pm$ 0.034 \\
\textbf{BA ($n$ = 1000, $m_{_{BA}}$ = 3)} & 0.089 $\pm$ 0.116 & 0.014 $\pm$ 0.024 & 0.089 $\pm$ 0.116 & 0.014 $\pm$ 0.024 & 0.082 $\pm$ 0.086 & 0.029 $\pm$ 0.031 \\
\textbf{BA ($n$ = 1000, $m_{_{BA}}$ = 4)} & 0.064 $\pm$ 0.068 & 0.017 $\pm$ 0.024 & 0.064 $\pm$ 0.068 & 0.017 $\pm$ 0.024 & 0.059 $\pm$ 0.052 & 0.028 $\pm$ 0.031 \\
\textbf{BA ($n$ = 1000, $m_{_{BA}}$ = 5)} & 0.050 $\pm$ 0.054 & 0.017 $\pm$ 0.025 & 0.050 $\pm$ 0.054 & 0.017 $\pm$ 0.025 & 0.047 $\pm$ 0.045 & 0.022 $\pm$ 0.033 \\
\textbf{BA ($n$ = 2000, $m_{_{BA}}$ = 2)} & 0.072 $\pm$ 0.164 & 0.009 $\pm$ 0.019 & 0.072 $\pm$ 0.164 & 0.009 $\pm$ 0.019 & 0.069 $\pm$ 0.112 & 0.023 $\pm$ 0.020 \\
\textbf{BA ($n$ = 2000, $m_{_{BA}}$ = 3)} & 0.058 $\pm$ 0.098 & 0.011 $\pm$ 0.017 & 0.058 $\pm$ 0.098 & 0.011 $\pm$ 0.017 & 0.054 $\pm$ 0.068 & 0.019 $\pm$ 0.024 \\
\bottomrule
\end{tabular}
\end{sidewaystable}

% Table S4
\begin{table}
\centering
\caption{Comparison of Steinerberger curvature with Forman-Ricci, Augmented Forman-Ricci, and Ollivier-Ricci curvature of vertices in real-world networks. Both Pearson and Spearman correlations are listed in this table up to three decimal places.}
\begin{tabular}{l c c c c c c}
\toprule
\toprule
\multirow{2}{*}{\textbf{Real-world Networks}} & \multicolumn{2}{c}{\textbf{Forman-Ricci}} & \multicolumn{2}{c}{\textbf{Augmented Forman-Ricci}} & \multicolumn{2}{c}{\textbf{Ollivier-Ricci}} \\
\cmidrule(lr){2-3}
\cmidrule(lr){4-5}
\cmidrule(lr){6-7}
 & \textbf{Pearson} & \textbf{Spearman} & \textbf{Pearson} & \textbf{Spearman} & \textbf{Pearson} & \textbf{Spearman}\\
\midrule
\textbf{Adjective-Noun adjacency} & 0.425 & 0.144 & 0.412 & 0.13 & 0.291 & 0.197 \\
\textbf{Chicago road} & 0.952 & 0.514 & 0.952 & 0.514 & 0.38 & 0.352 \\
\textbf{Contiguous US States} & 0.28 & 0.428 & 0.097 & 0.273 & 0.058 & 0.221 \\
\textbf{Dolphin} & 0.123 & 0.25 & -0.046 & 0.083 & -0.263 & -0.184 \\
\textbf{Email communication} & 0.036 & 0.077 & 0.04 & 0.077 & 0.03 & 0.095 \\
\textbf{Euro road} & 0.07 & 0.143 & 0.064 & 0.141 & 0.083 & 0.178 \\
\textbf{Facebook} & 0.969 & 0.013 & 0.969 & 0.012 & 0.584 & 0.115 \\
\textbf{Football} & 0.007 & 0.022 & 0.042 & 0.086 & -0.027 & 0.006 \\
\textbf{Hamsterster friendships} & 0.134 & 0.117 & 0.137 & 0.118 & 0.152 & 0.141 \\
\textbf{Human protein interactions} & 0.12 & 0.092 & 0.121 & 0.085 & 0.109 & 0.114 \\
\textbf{Jazz musicians} & 0.017 & 0.044 & 0.006 & 0.069 & -0.03 & -0.006 \\
\textbf{PDZ-domain interaction} & 0.532 & 0.246 & 0.531 & 0.239 & 0.385 & 0.198 \\
\textbf{Reactome} & -0.006 & 0.052 & 0.001 & 0.02 & 0.004 & -0.018 \\
\textbf{Route views} & -0.386 & 0.024 & -0.386 & 0.024 & -0.311 & 0.041 \\
\textbf{US Power Grid } & 0.028 & 0.105 & 0.035 & 0.097 & 0.035 & 0.121 \\
\textbf{Yeast protein interactions} & 0.312 & 0.181 & 0.31 & 0.139 & 0.242 & 0.199 \\
\textbf{Zachary karate club} & -0.028 & 0.082 & -0.015 & -0.104 & 0.074 & -0.051 \\
\textbf{Zebra} & -0.015 & 0.322 & 0.037 & -0.176 & 0.087 & -0.146 \\
\bottomrule
\end{tabular}
\end{table}

% Table S5
\begin{table}
\centering
\caption{Pearson and Spearman correlation along with standard deviation of resistance curvature with Forman-Ricci and Augmented Forman-Ricci curvature of vertices in the ER, WS, BA, and HGG model networks generated with different parameter values. The correlations are given up to three decimal places.}
\begin{tabular}{l@{\hspace{0.5cm}} c@{\hspace{0.4cm}} c@{\hspace{0.5cm}} c@{\hspace{0.4cm}} c@{\hspace{0.5cm}}}
\toprule
\toprule
\multirow{2}{*}{\textbf{Model Networks}} & \multicolumn{2}{c}{\textbf{Forman-Ricci}} & \multicolumn{2}{c}{\textbf{Augmented Forman-Ricci}} \\
\cmidrule(lr){2-3}
\cmidrule(lr){4-5}
 & \textbf{Pearson} & \textbf{Spearman} & \textbf{Pearson} & \textbf{Spearman} \\
\midrule
\textbf{ER ($n$ = 1000, $p_{_{ER}}$ = 0.003)} & 0.792 $\pm$ 0.012 & 0.792 $\pm$ 0.012 & 0.793 $\pm$ 0.012 & 0.792 $\pm$ 0.012 \\
\textbf{ER ($n$ = 1000, $p_{_{ER}}$ = 0.005)} & 0.831 $\pm$ 0.010 & 0.850 $\pm$ 0.009 & 0.832 $\pm$ 0.010 & 0.850 $\pm$ 0.009 \\
\textbf{ER ($n$ = 1000, $p_{_{ER}}$ = 0.007)} & 0.869 $\pm$ 0.008 & 0.881 $\pm$ 0.007 & 0.870 $\pm$ 0.008 & 0.882 $\pm$ 0.007 \\
\textbf{ER ($n$ = 1000, $p_{_{ER}}$ = 0.01)} & 0.906 $\pm$ 0.005 & 0.911 $\pm$ 0.005 & 0.907 $\pm$ 0.005 & 0.912 $\pm$ 0.005 \\
\textbf{ER ($n$ = 2000, $p_{_{ER}}$ = 0.0015)} & 0.793 $\pm$ 0.009 & 0.793 $\pm$ 0.009 & 0.793 $\pm$ 0.009 & 0.793 $\pm$ 0.009 \\
\textbf{ER ($n$ = 2000, $p_{_{ER}}$ = 0.0025)} & 0.831 $\pm$ 0.008 & 0.850 $\pm$ 0.007 & 0.832 $\pm$ 0.008 & 0.850 $\pm$ 0.007 \\
\bottomrule
\end{tabular}
\end{table}

% Table S6
\begin{table}
\centering
\caption{Pearson and Spearman correlation along with standard deviation of resistance curvature with Ollivier-Ricci curvature of vertices and Steinerberger curvature in the ER, WS, BA, and HGG model networks generated with different parameter values. The correlations are given up to three decimal places.}
\begin{tabular}{l@{\hspace{0.5cm}} c@{\hspace{0.4cm}} c@{\hspace{0.5cm}} c@{\hspace{0.4cm}} c@{\hspace{0.5cm}}}
\toprule
\toprule
\multirow{2}{*}{\textbf{Model Networks}} & \multicolumn{2}{c}{\textbf{Ollivier-Ricci}} & \multicolumn{2}{c}{\textbf{Steinerberger}} \\
\cmidrule(lr){2-3}
\cmidrule(lr){4-5}
 & \textbf{Pearson} & \textbf{Spearman} & \textbf{Pearson} & \textbf{Spearman} \\
 \midrule
\textbf{ER ($n$ = 1000, $p_{_{ER}}$ = 0.003)} & 0.785 $\pm$ 0.010 & 0.787 $\pm$ 0.011 & 0.142 $\pm$ 0.099 & 0.168 $\pm$ 0.104 \\
\textbf{ER ($n$ = 1000, $p_{_{ER}}$ = 0.005)} & 0.865 $\pm$ 0.009 & 0.872 $\pm$ 0.009 & 0.076 $\pm$ 0.058 & 0.072 $\pm$ 0.053 \\
\textbf{ER ($n$ = 1000, $p_{_{ER}}$ = 0.007)} & 0.918 $\pm$ 0.006 & 0.920 $\pm$ 0.006 & 0.041 $\pm$ 0.039 & 0.035 $\pm$ 0.037 \\
\textbf{ER ($n$ = 1000, $p_{_{ER}}$ = 0.01)} & 0.959 $\pm$ 0.003 & 0.957 $\pm$ 0.003 & 0.029 $\pm$ 0.036 & 0.027 $\pm$ 0.034 \\
\textbf{ER ($n$ = 2000, $p_{_{ER}}$ = 0.0015)} & 0.785 $\pm$ 0.008 & 0.786 $\pm$ 0.009 & 0.151 $\pm$ 0.094 & 0.177 $\pm$ 0.100 \\
\textbf{ER ($n$ = 2000, $p_{_{ER}}$ = 0.0025)} & 0.862 $\pm$ 0.007 & 0.867 $\pm$ 0.007 & 0.067 $\pm$ 0.048 & 0.064 $\pm$ 0.044 \\
\bottomrule
\end{tabular}
\end{table}

% Table S7
\begin{sidewaystable}
\centering
\caption{Pearson and Spearman correlation along with standard deviation of resistance curvature with Steinerberger curvature, Forman-Ricci, Augmented Forman-Ricci, and Ollivier-Ricci curvature of vertices in real-world networks. The correlations are given up to three decimal places.}
\begin{tabular}{l c c c c c c c c}
\toprule
\toprule
\multirow{2}{*}{\textbf{Real-world Networks}} & \multicolumn{2}{c}{\textbf{Steinerberger}} & \multicolumn{2}{c}{\textbf{Forman-Ricci}} & \multicolumn{2}{c}{\textbf{Augmented Forman-Ricci}} & \multicolumn{2}{c}{\textbf{Ollivier-Ricci }} \\
\cmidrule(lr){2-3}
\cmidrule(lr){4-5}
\cmidrule(lr){6-7}
\cmidrule(lr){8-9}
 & \textbf{Pearson} & \textbf{Spearman} & \textbf{Pearson} & \textbf{Spearman} & \textbf{Pearson} & \textbf{Spearman} & \textbf{Pearson} & \textbf{Spearman} \\
\midrule
\textbf{Adjective-Noun adjacency} & 0.438 & 0.245 & 0.905 & 0.825 & 0.913 & 0.824 & 0.876 & 0.733 \\
\textbf{Chicago road} & 1 & 0.28 & 0.952 & 0.405 & 0.952 & 0.405 & 0.38 & 0.212 \\
\textbf{Contiguous US States} & 0.466 & 0.488 & 0.783 & 0.808 & 0.651 & 0.625 & 0.584 & 0.5 \\
\textbf{Dolphin} & 0.162 & 0.218 & 0.809 & 0.862 & 0.671 & 0.656 & 0.135 & 0.134 \\
\textbf{Email communication} & 0.082 & 0.122 & 0.777 & 0.849 & 0.783 & 0.819 & 0.835 & 0.793 \\
\textbf{Euro road} & 0.155 & 0.252 & 0.719 & 0.539 & 0.728 & 0.534 & 0.774 & 0.627 \\
\bottomrule
\end{tabular}
\end{sidewaystable}

% Table S8
\begin{table}
\centering
\caption{Pearson and Spearman correlation along with standard deviation of scalar sectional curvature with resistance curvature and Steinerberger curvature in the model networks generated with different parameter values. The correlations are given up to 3 decimal places.}
\begin{tabular}{l@{\hspace{0.5cm}} c@{\hspace{0.4cm}} c@{\hspace{0.5cm}} c@{\hspace{0.4cm}} c@{\hspace{0.5cm}}}
\toprule
\toprule
\multirow{2}{*}{\textbf{Model Networks}} & \multicolumn{2}{c}{\textbf{Resistance}} & \multicolumn{2}{c}{\textbf{Steinerberger}} \\
 \cmidrule(lr){2-3}
\cmidrule(lr){4-5}
 & \textbf{Pearson} & \textbf{Spearman} & \textbf{Pearson} & \textbf{Spearman} \\
\midrule
\textbf{ER ($n$ = 1000, $p_{_{ER}}$ = 0.003)} & -0.530 $\pm$ 0.025 & -0.581 $\pm$ 0.028 & -0.115 $\pm$ 0.077 & -0.126 $\pm$ 0.079 \\
\textbf{ER ($n$ = 1000, $p_{_{ER}}$ = 0.005)} & -0.641 $\pm$ 0.035 & -0.655 $\pm$ 0.025 & -0.025 $\pm$ 0.031 & -0.021 $\pm$ 0.032 \\
\textbf{ER ($n$ = 1000, $p_{_{ER}}$ = 0.007)} & -0.647 $\pm$ 0.024 & -0.746 $\pm$ 0.015 & -0.043 $\pm$ 0.034 & -0.032 $\pm$ 0.024 \\
\textbf{ER ($n$ = 1000, $p_{_{ER}}$ = 0.01)} & 0.285 $\pm$ 0.101 & 0.249 $\pm$ 0.096 & 0.010 $\pm$ 0.017 & 0.012 $\pm$ 0.017 \\
\textbf{WS ($n$ = 1000, $K_{_{WS}}$ = 2)} & -0.021 $\pm$ 0.024 & -0.020 $\pm$ 0.033 & -0.024 $\pm$ 0.020 & -0.041 $\pm$ 0.043 \\
\textbf{WS ($n$ = 1000, $K_{_{WS}}$ = 4)} & -0.591 $\pm$ 0.013 & -0.571 $\pm$ 0.016 & -0.022 $\pm$ 0.024 & -0.023 $\pm$ 0.022 \\
\bottomrule
\end{tabular}
\end{table}

% Table S9
\begin{table}
\centering
\caption{Pearson and Spearman correlations along with standard deviation of scalar sectional curvature with resistance curvature and Steinerberger curvature in the real-world networks. The correlations are given up to 3 decimal places.}
\begin{tabular}{ l  c  c  c  c  }
\toprule
\toprule
\multirow{2}{*}{\textbf{Real-world Networks}} & \multicolumn{2}{c}{\textbf{Resistance}} &\multicolumn{2}{c}{\textbf{Steinerberger}}     \\
\cmidrule(lr){2-3}
\cmidrule(lr){4-5}&  \textbf{Pearson}& \textbf{Spearman} & \multicolumn{1}{c}{\textbf{Pearson}} & \multicolumn{1}{c}{\textbf{Spearman}}        \\
\midrule
\textbf{Adjective-Noun adjacency} &  -0.139& -0.125&-0.052 & -0.086 \\
\textbf{Chicago road} &  0 & 0 & 0 & 0 \\
\textbf{Contiguous US States} &  -0.41& -0.441&-0.304 & -0.354 \\
\textbf{Dolphin} &  -0.625& -0.71&-0.227 & -0.254 \\
\textbf{Email communication} &  -0.507& -0.548&-0.077 & -0.102 \\
\textbf{Euro road} &  -0.116& -0.056&-0.014 & -0.001 \\
\textbf{Football} &  -0.365& -0.32&0.034 & -0.02 \\
\textbf{Jazz musicians} &  -0.285& -0.255&-0.012 & 0.000 \\
\textbf{PDZ-domain interaction} &  -0.07& -0.144&-0.212 & -0.164 \\
\textbf{Yeast protein interactions} &  -0.23& -0.287&-0.091 & -0.098 \\
\textbf{Zachary karate club} &  -0.626& -0.603&0.047 & -0.135 \\
\textbf{Zebra} &  -0.452& -0.331&-0.103 & -0.419 \\
\bottomrule
\end{tabular}
\end{table}

% Table S10
\begin{table}
\caption{Pearson and Spearman correlation along with standard deviation of Steinerberger curvature with degree and clustering coefficient in the model networks generated with different parameter values. The correlations are given up to 3 decimal places.}
\begin{tabular}{l@{\hspace{0.5cm}} c@{\hspace{0.4cm}} c@{\hspace{0.5cm}} c@{\hspace{0.4cm}} c@{\hspace{0.5cm}} }
\toprule
\toprule
\multirow{2}{*}{\textbf{Model Networks}} & \multicolumn{2}{c}{\textbf{Degree}} & \multicolumn{2}{c}{\textbf{Clustering coefficient}} \\
\cmidrule(lr){2-3}
\cmidrule(lr){4-5}
& \multicolumn{1}{c}{\textbf{Pearson}} & \multicolumn{1}{c}{\textbf{Spearman}} & \multicolumn{1}{c}{\textbf{Pearson}} & \multicolumn{1}{c}{\textbf{Spearman}} \\
\midrule
\textbf{ER ($n$ = 1000, $p_{_{ER}}$ = 0.003)} & -0.114 $\pm$ 0.082 & -0.147 $\pm$ 0.094 & -0.000 $\pm$ 0.028 & -0.006 $\pm$ 0.027 \\
\textbf{ER ($n$ = 1000, $p_{_{ER}}$ = 0.005)} & -0.057 $\pm$ 0.046 & -0.058 $\pm$ 0.044 & -0.005 $\pm$ 0.031 & -0.009 $\pm$ 0.023 \\
\textbf{ER ($n$ = 1000, $p_{_{ER}}$ = 0.007)} & -0.031 $\pm$ 0.032 & -0.027 $\pm$ 0.032 & -0.001 $\pm$ 0.027 & -0.003 $\pm$ 0.027 \\
\textbf{ER ($n$ = 1000, $p_{_{ER}}$ = 0.01)} & -0.026 $\pm$ 0.034 & -0.024 $\pm$ 0.032 & -0.003 $\pm$ 0.030 & -0.008 $\pm$ 0.029 \\
\textbf{ER ($n$ = 2000, $p_{_{ER}}$ = 0.0015)} & -0.123 $\pm$ 0.076 & -0.155 $\pm$ 0.089 & -0.008 $\pm$ 0.024 & -0.008 $\pm$ 0.016 \\
\textbf{ER ($n$ = 2000, $p_{_{ER}}$ = 0.0025)} & -0.047 $\pm$ 0.036 & -0.051 $\pm$ 0.035 & -0.002 $\pm$ 0.022 & -0.005 $\pm$ 0.019 \\
\bottomrule
\end{tabular}
\end{table}

% Table S11
\begin{sidewaystable}
\centering
\caption{Pearson and Spearman correlation along with standard deviation of Steinerberger curvature with betweenness centrality, eigenvector centrality, and closeness centrality in the ER and WS model networks generated with different parameter values. The correlations are given up to 3 decimal places.}
\begin{tabular}{l@{\hspace{0.5cm}} c@{\hspace{0.4cm}} c@{\hspace{0.5cm}} c@{\hspace{0.4cm}} c@{\hspace{0.5cm}} c@{\hspace{0.4cm}} c@{\hspace{0.5cm}}}
\toprule
\toprule
\multirow{2}{*}{\textbf{Model Networks}} & \multicolumn{2}{c}{\textbf{Betweenness centrality}} & \multicolumn{2}{c}{\textbf{Eigenvector centrality}} & \multicolumn{2}{c}{\textbf{Closeness centrality}} \\
 \cmidrule(lr){2-3}
\cmidrule(lr){4-5}
\cmidrule(lr){6-7}
& \multicolumn{1}{c}{\textbf{Pearson}} & \multicolumn{1}{c}{\textbf{Spearman}} & \multicolumn{1}{c}{\textbf{Pearson}} & \multicolumn{1}{c}{\textbf{Spearman}} & \multicolumn{1}{c}{\textbf{Pearson}} & \multicolumn{1}{c}{\textbf{Spearman}} \\
\midrule
\textbf{ER ($n$ = 1000, $p_{_{ER}}$ = 0.003)} & -0.080 $\pm$ 0.064 & -0.142 $\pm$ 0.087 & -0.055 $\pm$ 0.046 & -0.095 $\pm$ 0.059 & -0.084 $\pm$ 0.056 & -0.099 $\pm$ 0.063 \\
\textbf{ER ($n$ = 1000, $p_{_{ER}}$ = 0.005)} & -0.048 $\pm$ 0.042 & -0.062 $\pm$ 0.045 & -0.039 $\pm$ 0.033 & -0.048 $\pm$ 0.036 & -0.057 $\pm$ 0.040 & -0.048 $\pm$ 0.036 \\
\textbf{ER ($n$ = 1000, $p_{_{ER}}$ = 0.007)} & -0.027 $\pm$ 0.033 & -0.028 $\pm$ 0.033 & -0.023 $\pm$ 0.026 & -0.021 $\pm$ 0.027 & -0.032 $\pm$ 0.027 & -0.021 $\pm$ 0.028 \\
\textbf{ER ($n$ = 1000, $p_{_{ER}}$ = 0.01)} & -0.023 $\pm$ 0.034 & -0.024 $\pm$ 0.033 & -0.023 $\pm$ 0.029 & -0.021 $\pm$ 0.029 & -0.026 $\pm$ 0.031 & -0.021 $\pm$ 0.030 \\
\textbf{ER ($n$ = 2000, $p_{_{ER}}$ = 0.0015)} & -0.085 $\pm$ 0.054 & -0.147 $\pm$ 0.082 & -0.060 $\pm$ 0.038 & -0.098 $\pm$ 0.056 & -0.089 $\pm$ 0.052 & -0.102 $\pm$ 0.058 \\
\textbf{ER ($n$ = 2000, $p_{_{ER}}$ = 0.0025)} & -0.037 $\pm$ 0.031 & -0.054 $\pm$ 0.036 & -0.031 $\pm$ 0.025 & -0.042 $\pm$ 0.027 & -0.049 $\pm$ 0.033 & -0.042 $\pm$ 0.027 \\
\bottomrule
\end{tabular}
\end{sidewaystable}

% Table S12
\begin{sidewaystable}
\centering
\caption{Pearson and Spearman correlation along with standard deviation of Steinerberger curvature with betweenness centrality, eigenvector centrality, and closeness centrality in the BA and HGG model networks generated with different parameter values. The correlations are given up to 3 decimal places.}
\begin{tabular}{l@{\hspace{0.5cm}} c@{\hspace{0.4cm}} c@{\hspace{0.5cm}} c@{\hspace{0.4cm}} c@{\hspace{0.5cm}} c@{\hspace{0.4cm}} c@{\hspace{0.5cm}}}
\toprule
\toprule
\multirow{2}{*}{\textbf{Model Networks}} & \multicolumn{2}{c}{\textbf{Betweenness centrality}} & \multicolumn{2}{c}{\textbf{Eigenvector centrality}} & \multicolumn{2}{c}{\textbf{Closeness centrality}} \\
 \cmidrule(lr){2-3}
\cmidrule(lr){4-5}
\cmidrule(lr){6-7}
& \multicolumn{1}{c}{\textbf{Pearson}} & \multicolumn{1}{c}{\textbf{Spearman}} & \multicolumn{1}{c}{\textbf{Pearson}} & \multicolumn{1}{c}{\textbf{Spearman}} & \multicolumn{1}{c}{\textbf{Pearson}} & \multicolumn{1}{c}{\textbf{Spearman}} \\
\midrule
\textbf{BA ($n$ = 1000, $m_{_{BA}}$ = 2)} & -0.052 $\pm$ 0.158 & -0.022 $\pm$ 0.035 & -0.049 $\pm$ 0.119 & -0.007 $\pm$ 0.027 & -0.034 $\pm$ 0.030 & -0.010 $\pm$ 0.029 \\
\textbf{BA ($n$ = 1000, $m_{_{BA}}$ = 3)} & -0.089 $\pm$ 0.107 & -0.021 $\pm$ 0.028 & -0.077 $\pm$ 0.081 & -0.010 $\pm$ 0.021 & -0.043 $\pm$ 0.031 & -0.011 $\pm$ 0.021 \\
\textbf{BA ($n$ = 1000, $m_{_{BA}}$ = 4)} & -0.065 $\pm$ 0.064 & -0.023 $\pm$ 0.031 & -0.057 $\pm$ 0.049 & -0.012 $\pm$ 0.019 & -0.037 $\pm$ 0.027 & -0.013 $\pm$ 0.020 \\
\textbf{BA ($n$ = 1000, $m_{_{BA}}$ = 5)} & -0.050 $\pm$ 0.052 & -0.019 $\pm$ 0.033 & -0.046 $\pm$ 0.042 & -0.014 $\pm$ 0.019 & -0.033 $\pm$ 0.029 & -0.015 $\pm$ 0.020 \\
\textbf{BA ($n$ = 2000, $m_{_{BA}}$ = 2)} & -0.075 $\pm$ 0.148 & -0.016 $\pm$ 0.020 & -0.060 $\pm$ 0.119 & -0.004 $\pm$ 0.016 & -0.028 $\pm$ 0.022 & -0.005 $\pm$ 0.017 \\
\textbf{BA ($n$ = 2000, $m_{_{BA}}$ = 3)} & -0.059 $\pm$ 0.089 & -0.015 $\pm$ 0.021 & -0.051 $\pm$ 0.070 & -0.007 $\pm$ 0.015 & -0.025 $\pm$ 0.020 & -0.008 $\pm$ 0.016 \\
\bottomrule
\end{tabular}
\end{sidewaystable}

% Table S13
\begin{sidewaystable}
\centering
\caption{Comparison of Steinerberger curvature with degree, clustering coefficient, betweenness centrality, eigenvector centrality, and closeness centrality in real-world networks. Both Spearman and Pearson correlations are given in this table up to 3 decimal places.}
\begin{tabular}{ l  c  c  c  c  c  c  c  c  c  c }
\toprule
\toprule
\multirow{2}{*}{\textbf{Real-world Networks}} & \multicolumn{2}{c}{\textbf{Degree}} & \multicolumn{2}{c}{\textbf{Clustering coefficient}} & \multicolumn{2}{c}{\textbf{Betweenness centrality}} & \multicolumn{2}{c}{\textbf{Eigenvector centrality}} & \multicolumn{2}{c}{\textbf{Closeness centrality}} \\
 \cmidrule(lr){2-3}
\cmidrule(lr){4-5}
\cmidrule(lr){6-7}
\cmidrule(lr){8-9}
\cmidrule(lr){10-11}
 & \multicolumn{1}{c}{\textbf{Pearson}} & \multicolumn{1}{c}{\textbf{Spearman}} & \multicolumn{1}{c}{\textbf{Pearson}} & \multicolumn{1}{c}{\textbf{Spearman}} & \multicolumn{1}{c}{\textbf{Pearson}} & \multicolumn{1}{c}{\textbf{Spearman}} & \multicolumn{1}{c}{\textbf{Pearson}} & \multicolumn{1}{c}{\textbf{Spearman}} & \multicolumn{1}{c}{\textbf{Pearson}} & \multicolumn{1}{c}{\textbf{Spearman}} \\
\midrule
\textbf{Adjective-Noun adjacency} & -0.403 & -0.194 & 0.083 & -0.016 & -0.427 & -0.228 & -0.362 & -0.13 & -0.278 & -0.131 \\
\textbf{Chicago road} & -1 & -0.601 & 0 & 0 & -0.405 & -0.6 & -0.713 & -0.029 & -0.721 & -0.019 \\
\textbf{Contiguous US States} & -0.398 & -0.468 & 0.302 & 0.261 & -0.196 & -0.432 & -0.184 & -0.23 & -0.22 & -0.259 \\
\textbf{Dolphin} & -0.147 & -0.231 & -0.305 & -0.379 & 0.141 & -0.148 & -0.102 & -0.109 & -0.05 & -0.109 \\
\textbf{Email communication} & -0.038 & -0.089 & -0.017 & -0.062 & -0.042 & -0.117 & -0.039 & -0.063 & -0.058 & -0.072 \\
\textbf{Euro road} & -0.118 & -0.236 & -0.023 & -0.058 & -0.046 & -0.232 & -0.001 & 0.001 & -0.019 & -0.012 \\
\bottomrule
\end{tabular}
\label{tab:realGraphCorr}
\end{sidewaystable}

% Table S14
\begin{table}
\caption{Pearson and Spearman correlation along with standard deviation of resistance curvature with degree and clustering coefficient in the model networks generated with different parameter values. The correlations are given up to 3 decimal places.}
\begin{tabular}{l@{\hspace{0.5cm}} c@{\hspace{0.4cm}} c@{\hspace{0.5cm}} c@{\hspace{0.4cm}} c@{\hspace{0.5cm}} }
\toprule
\toprule
\multirow{2}{*}{\textbf{Model Networks}} & \multicolumn{2}{c}{\textbf{Degree}} & \multicolumn{2}{c}{\textbf{Clustering coefficient}} \\
\cmidrule(lr){2-3}
\cmidrule(lr){4-5}
 & \textbf{Pearson} & \textbf{Spearman} & \textbf{Pearson} & \textbf{Spearman} \\
\midrule
\textbf{ER ($n$ = 1000, $p_{_{ER}}$ = 0.003)} & -0.925 $\pm$ 0.005 & -0.933 $\pm$ 0.005 & -0.005 $\pm$ 0.030 & -0.083 $\pm$ 0.030 \\
\textbf{ER ($n$ = 1000, $p_{_{ER}}$ = 0.005)} & -0.918 $\pm$ 0.006 & -0.928 $\pm$ 0.006 & 0.006 $\pm$ 0.029 & -0.150 $\pm$ 0.028 \\
\textbf{ER ($n$ = 1000, $p_{_{ER}}$ = 0.007)} & -0.933 $\pm$ 0.005 & -0.938 $\pm$ 0.005 & 0.011 $\pm$ 0.033 & -0.198 $\pm$ 0.024 \\
\textbf{ER ($n$ = 1000, $p_{_{ER}}$ = 0.01)} & -0.953 $\pm$ 0.003 & -0.952 $\pm$ 0.003 & 0.027 $\pm$ 0.033 & -0.192 $\pm$ 0.029 \\
\textbf{ER ($n$ = 2000, $p_{_{ER}}$ = 0.0015)} & -0.925 $\pm$ 0.004 & -0.933 $\pm$ 0.004 & -0.002 $\pm$ 0.018 & -0.058 $\pm$ 0.020 \\
\textbf{ER ($n$ = 2000, $p_{_{ER}}$ = 0.0025)} & -0.919 $\pm$ 0.004 & -0.928 $\pm$ 0.004 & 0.005 $\pm$ 0.023 & -0.106 $\pm$ 0.021 \\
\bottomrule
\end{tabular}
\end{table}

% Table S15
\begin{sidewaystable}
\centering
\caption{Pearson and Spearman correlation along with standard deviation of resistance curvature with betweenness centrality, eigenvector centrality, and closeness centrality in the ER and WS model networks generated with different parameter values. The correlations are given up to 3 decimal places.}
\begin{tabular}{l@{\hspace{0.5cm}} c@{\hspace{0.4cm}} c@{\hspace{0.5cm}} c@{\hspace{0.4cm}} c@{\hspace{0.5cm}} c@{\hspace{0.4cm}} c@{\hspace{0.5cm}}}
\toprule
\toprule
\multirow{2}{*}{\textbf{Model Networks}} & \multicolumn{2}{c}{\textbf{Betweenness centrality}} & \multicolumn{2}{c}{\textbf{Eigenvector centrality}} & \multicolumn{2}{c}{\textbf{Closeness centrality}} \\
 \cmidrule(lr){2-3}
\cmidrule(lr){4-5}
\cmidrule(lr){6-7}
 & \textbf{Pearson} & \textbf{Spearman} & \textbf{Pearson} & \textbf{Spearman} & \textbf{Pearson} & \textbf{Spearman} \\
\midrule
\textbf{ER ($n$ = 1000, $p_{_{ER}}$ = 0.003)} & -0.783 $\pm$ 0.014 & -0.868 $\pm$ 0.008 & -0.536 $\pm$ 0.038 & -0.567 $\pm$ 0.024 & -0.574 $\pm$ 0.016 & -0.598 $\pm$ 0.019 \\
\textbf{ER ($n$ = 1000, $p_{_{ER}}$ = 0.005)} & -0.859 $\pm$ 0.009 & -0.904 $\pm$ 0.005 & -0.684 $\pm$ 0.017 & -0.697 $\pm$ 0.016 & -0.692 $\pm$ 0.013 & -0.712 $\pm$ 0.013 \\
\textbf{ER ($n$ = 1000, $p_{_{ER}}$ = 0.007)} & -0.901 $\pm$ 0.006 & -0.927 $\pm$ 0.004 & -0.759 $\pm$ 0.013 & -0.761 $\pm$ 0.013 & -0.759 $\pm$ 0.011 & -0.771 $\pm$ 0.011 \\
\textbf{ER ($n$ = 1000, $p_{_{ER}}$ = 0.01)} & -0.936 $\pm$ 0.003 & -0.952 $\pm$ 0.003 & -0.824 $\pm$ 0.008 & -0.818 $\pm$ 0.009 & -0.826 $\pm$ 0.007 & -0.826 $\pm$ 0.008 \\
\textbf{ER ($n$ = 2000, $p_{_{ER}}$ = 0.0015)} & -0.762 $\pm$ 0.010 & -0.856 $\pm$ 0.006 & -0.524 $\pm$ 0.042 & -0.565 $\pm$ 0.018 & -0.567 $\pm$ 0.013 & -0.592 $\pm$ 0.014 \\
\textbf{ER ($n$ = 2000, $p_{_{ER}}$ = 0.0025)} & -0.843 $\pm$ 0.007 & -0.892 $\pm$ 0.004 & -0.682 $\pm$ 0.013 & -0.697 $\pm$ 0.013 & -0.685 $\pm$ 0.010 & -0.708 $\pm$ 0.012 \\
\bottomrule
\end{tabular}
\label{tab:realGraphCorr}
\end{sidewaystable}

% Table S16
\begin{sidewaystable}
\centering
\caption{Pearson and Spearman correlation along with standard deviation of resistance curvature with betweenness centrality, eigenvector centrality, and closeness centrality in the BA and HGG model networks generated with different parameter values. The correlations are given up to 3 decimal places.}
\begin{tabular}{l@{\hspace{0.5cm}} c@{\hspace{0.4cm}} c@{\hspace{0.5cm}} c@{\hspace{0.4cm}} c@{\hspace{0.5cm}} c@{\hspace{0.4cm}} c@{\hspace{0.5cm}}}
\toprule
\toprule
\multirow{2}{*}{\textbf{Model Networks}} & \multicolumn{2}{c}{\textbf{Betweenness centrality}} & \multicolumn{2}{c}{\textbf{Eigenvector centrality}} & \multicolumn{2}{c}{\textbf{Closeness centrality}} \\
 \cmidrule(lr){2-3}
\cmidrule(lr){4-5}
\cmidrule(lr){6-7}
& \multicolumn{1}{c}{\textbf{Pearson}} & \multicolumn{1}{c}{\textbf{Spearman}} & \multicolumn{1}{c}{\textbf{Pearson}} & \multicolumn{1}{c}{\textbf{Spearman}} & \multicolumn{1}{c}{\textbf{Pearson}} & \multicolumn{1}{c}{\textbf{Spearman}} \\
\midrule
\textbf{BA ($n$ = 1000, $m_{_{BA}}$ = 2)} & -0.919 $\pm$ 0.017 & -0.704 $\pm$ 0.016 & -0.757 $\pm$ 0.021 & -0.174 $\pm$ 0.033 & -0.512 $\pm$ 0.034 & -0.194 $\pm$ 0.034 \\
\textbf{BA ($n$ = 1000, $m_{_{BA}}$ = 3)} & -0.920 $\pm$ 0.012 & -0.777 $\pm$ 0.012 & -0.834 $\pm$ 0.013 & -0.295 $\pm$ 0.026 & -0.614 $\pm$ 0.028 & -0.313 $\pm$ 0.027 \\
\textbf{BA ($n$ = 1000, $m_{_{BA}}$ = 4)} & -0.921 $\pm$ 0.008 & -0.825 $\pm$ 0.011 & -0.871 $\pm$ 0.008 & -0.383 $\pm$ 0.022 & -0.680 $\pm$ 0.024 & -0.397 $\pm$ 0.024 \\
\textbf{BA ($n$ = 1000, $m_{_{BA}}$ = 5)} & -0.924 $\pm$ 0.007 & -0.862 $\pm$ 0.010 & -0.894 $\pm$ 0.007 & -0.451 $\pm$ 0.024 & -0.733 $\pm$ 0.017 & -0.463 $\pm$ 0.026 \\
\textbf{BA ($n$ = 2000, $m_{_{BA}}$ = 2)} & -0.904 $\pm$ 0.018 & -0.674 $\pm$ 0.013 & -0.720 $\pm$ 0.026 & -0.162 $\pm$ 0.022 & -0.461 $\pm$ 0.032 & -0.183 $\pm$ 0.023 \\
\textbf{BA ($n$ = 2000, $m_{_{BA}}$ = 3)} & -0.906 $\pm$ 0.011 & -0.735 $\pm$ 0.010 & -0.811 $\pm$ 0.015 & -0.267 $\pm$ 0.019 & -0.556 $\pm$ 0.026 & -0.287 $\pm$ 0.020 \\
\bottomrule
\end{tabular}
\label{tab:realGraphCorr}
\end{sidewaystable}

% Table S17
\begin{sidewaystable}
\centering
\caption{Comparison of resistance curvature with degree, clustering coefficient, betweenness centrality, eigenvector centrality, and closeness centrality in real-world networks. Both Pearson and Spearman correlations are given in this table.}
\begin{tabular}{ l  c  c  c  c  c  c  c  c  c  c }
\toprule
\toprule
\multirow{2}{*}{\textbf{Real-world Networks}} & \multicolumn{2}{c}{\textbf{Degree}} & \multicolumn{2}{c}{\textbf{Clustering coefficient}} & \multicolumn{2}{c}{\textbf{Betweenness centrality}} & \multicolumn{2}{c}{\textbf{Eigenvector centrality}} & \multicolumn{2}{c}{\textbf{Closeness centrality}} \\
\cmidrule(lr){2-3}
\cmidrule(lr){4-5}
\cmidrule(lr){6-7}
\cmidrule(lr){8-9}
\cmidrule(lr){10-11}
 & \textbf{Pearson} & \textbf{Spearman} & \textbf{Pearson} & \textbf{Spearman} & \textbf{Pearson} & \textbf{Spearman} & \textbf{Pearson} & \textbf{Spearman} & \textbf{Pearson} & \textbf{Spearman} \\
\midrule
\textbf{Adjective-Noun adjacency} & -0.95 & -0.921 & 0.067 & -0.184 & -0.956 & -0.962 & -0.851 & -0.763 & -0.725 & -0.751 \\
\textbf{Chicago road} & -1 & -0.604 & 0 & 0 & -0.405 & -0.603 & -0.713 & -0.322 & -0.721 & -0.312 \\
\textbf{Contiguous US States} & -0.895 & -0.897 & 0.652 & 0.719 & -0.753 & -0.858 & -0.528 & -0.487 & -0.51 & -0.492 \\
\textbf{Dolphin} & -0.883 & -0.928 & -0.24 & -0.335 & -0.634 & -0.854 & -0.486 & -0.469 & -0.527 & -0.59 \\
\textbf{Email communication} & -0.87 & -0.921 & 0.139 & -0.208 & -0.857 & -0.948 & -0.642 & -0.728 & -0.692 & -0.766 \\
\textbf{Euro road} & -0.913 & -0.875 & -0.031 & -0.239 & -0.406 & -0.686 & -0.176 & -0.046 & -0.145 & -0.069 \\
\bottomrule
\end{tabular}
\end{sidewaystable}

\section{SUPPLEMENTARY FIGURES}

% Figure S1
\begin{figure}[ht]
    \centering
    \includegraphics{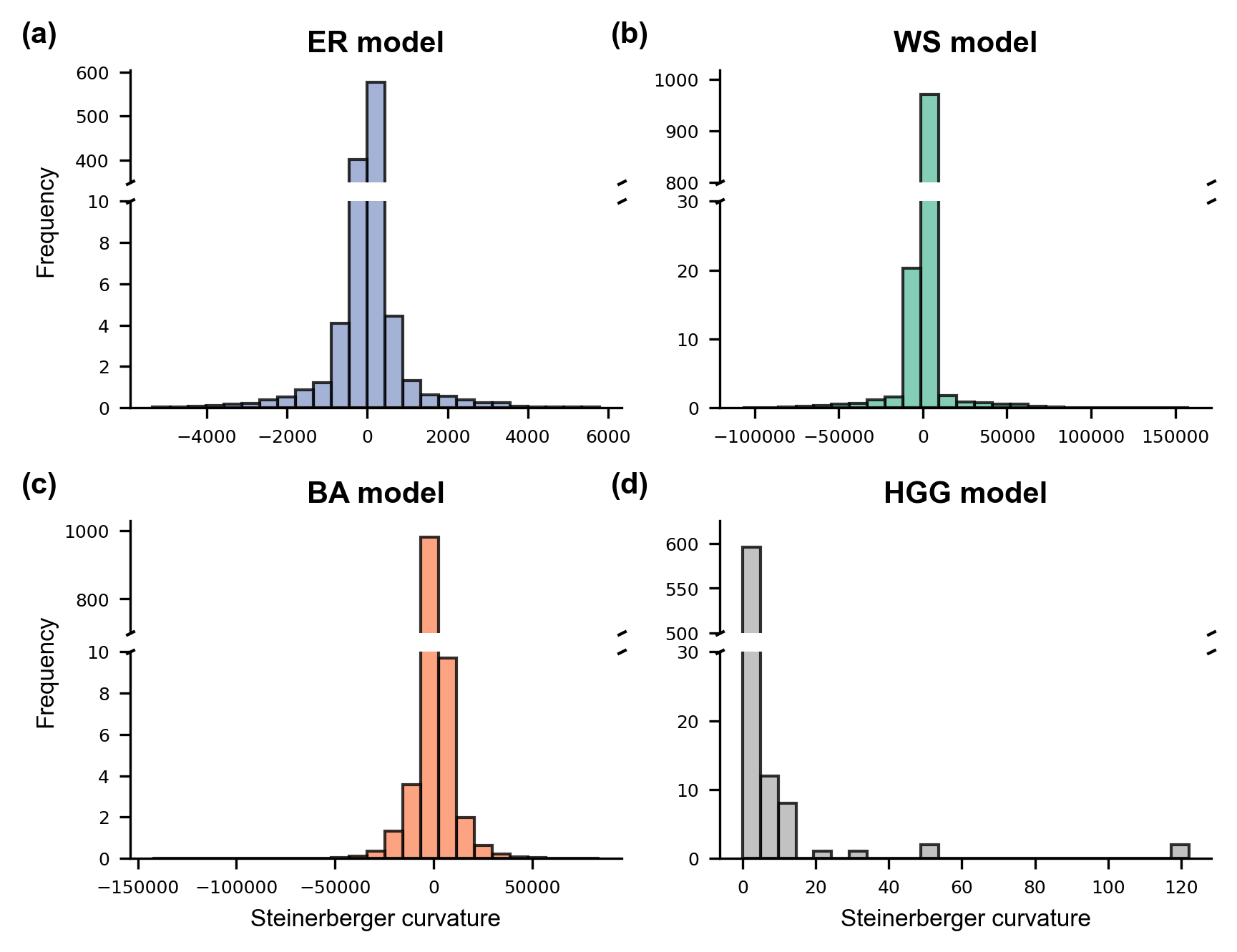}
    \caption{Distribution of Steinerberger curvatures across 100 independently generated model networks for each parameter combination, with $n=1000$ vertices: (a) ER model with $p_{\mathrm{ER}}=0.005$, (b) WS model with $k_{\mathrm{WS}}=6$, (c) BA model with $m_{\mathrm{BA}}=3$, and (d) HGG model with $k_{\mathrm{HGG}}=3$. A break in the y-axis is applied to accommodate the wide range of the frequency distributions.}
    % \label{fig:SBC_distribution_ModelSI}
\end{figure}

% Figure S2
\begin{figure}
    \centering
    \includegraphics{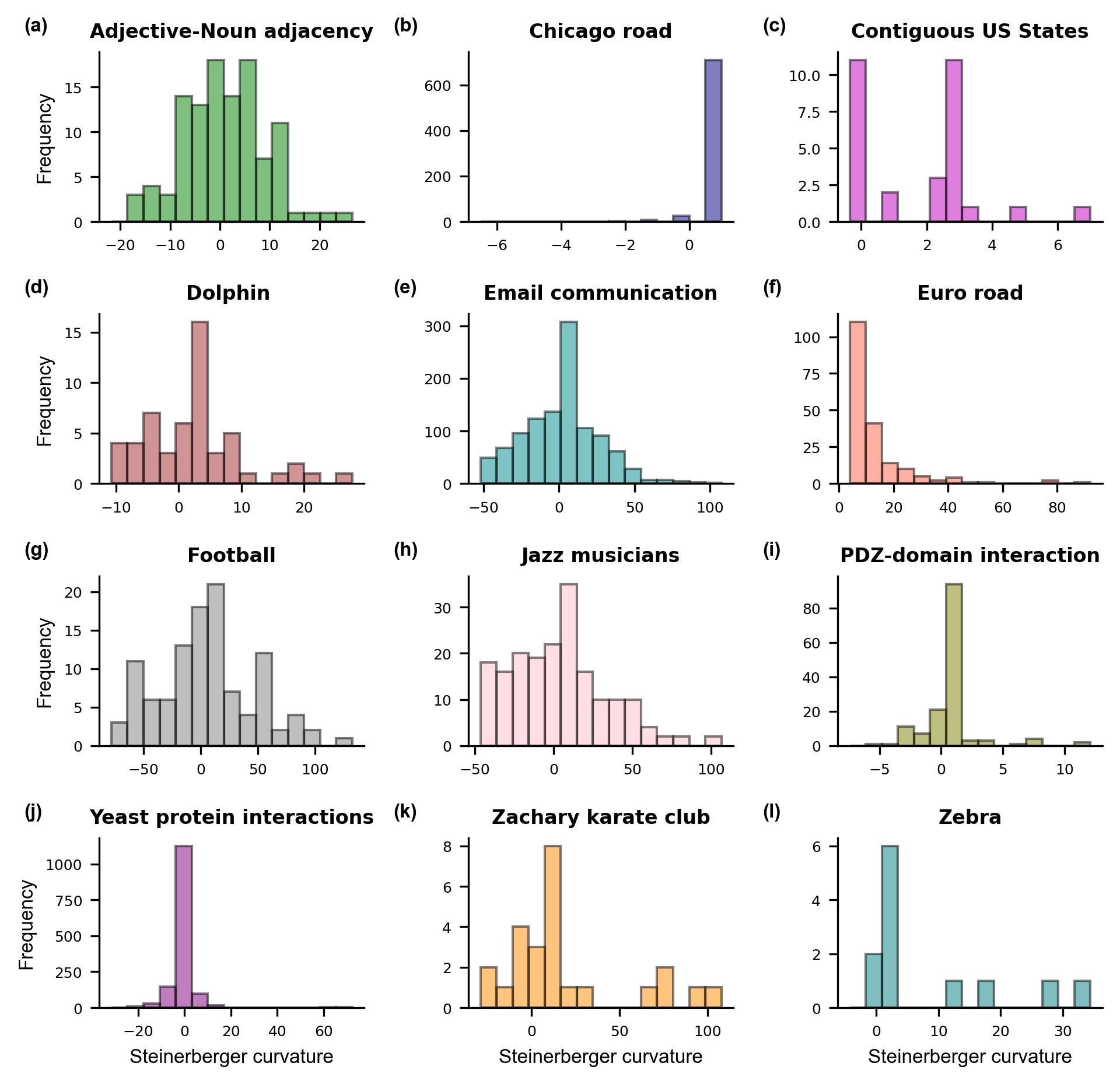}
    \caption{Distribution of Steinerberger curvature for different real-world networks. (a) Adjective-Noun Adjacency, (b) Chicago Road, (c) Contiguous US States, (d) Dolphin, (e) Email Communication, (f) Euro Road, (g) Football, (h) Jazz Musicians, (i) PDZ-Domain Interaction, (j) Yeast Protein Interactions, (k) Zachary Karate Club, and (l) Zebra networks.}
    % \label{fig:SBC_distribution_RealSI}
\end{figure}

% Figure S3
\begin{figure}
    \centering
    \includegraphics{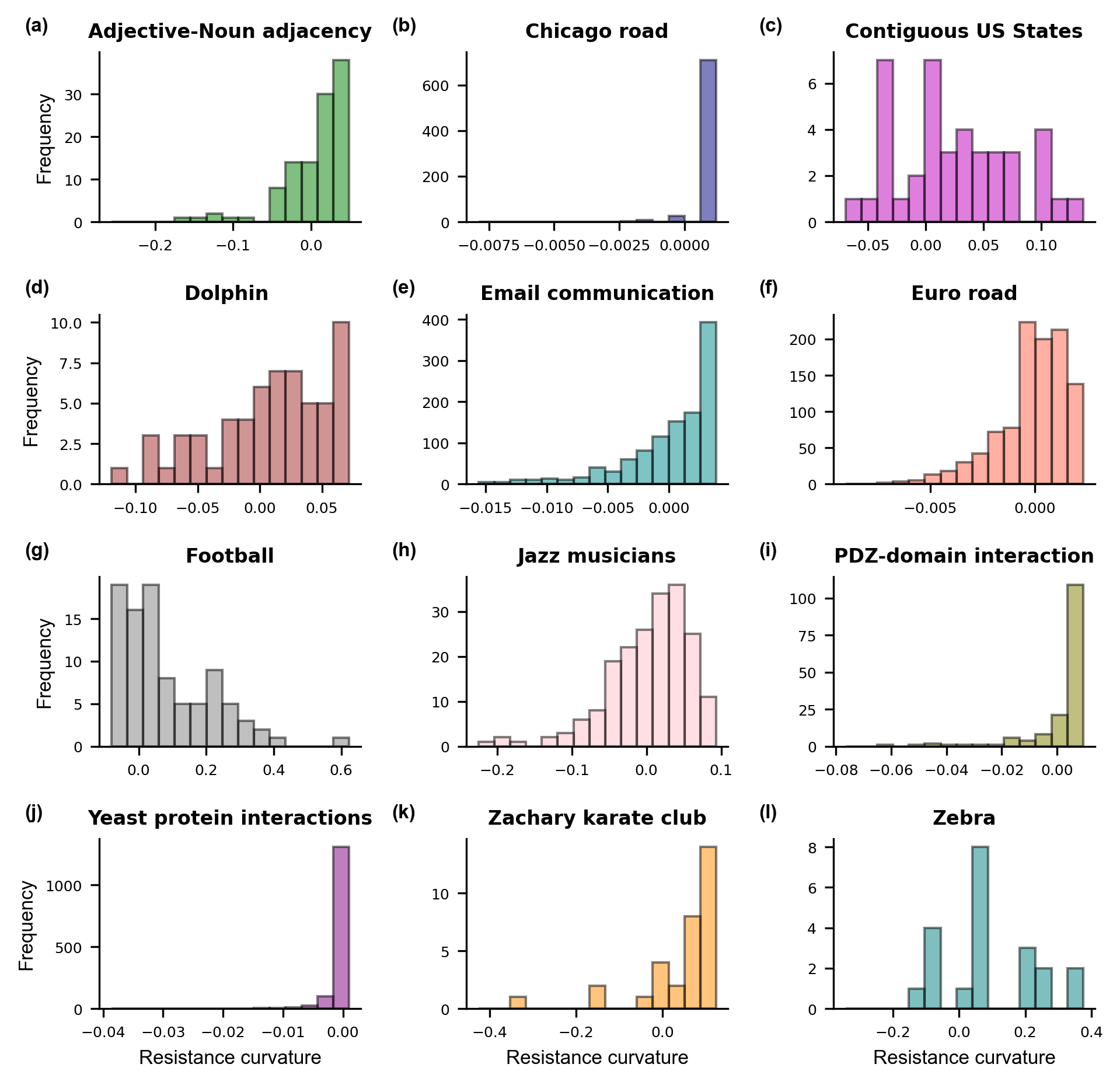}
    \caption{Distribution of resistance curvature for different real-world networks. (a) Adjective-Noun Adjacency, (b) Chicago Road, (c) Contiguous US States, (d) Dolphin, (e) Email Communication, (f) Euro Road, (g) Football, (h) Jazz Musicians, (i) PDZ-Domain Interaction, (j) Yeast Protein Interactions, (k) Zachary Karate Club, and (l) Zebra networks.}
    % \label{fig:SBC_distribution_RealSI}
\end{figure}

\end{document}